\documentclass[11pt]{article}
\usepackage{a4}

\newcommand{\cD}{{\cal D}}

\newcommand{\dmu}{{\dot \mu}}
\newcommand{\dnu}{{\dot \nu}}
\newcommand{\drho}{{\dot \rho}}

\def \IIA {$\mbox{I\hspace{-.1em}IA\hspace{.14em}}$}
\def \IIB {$\mbox{I\hspace{-.1em}IB\hspace{.14em}}$}

\newcommand{\ul}{\underline}
\newcommand{\wt}{\widetilde}
\newcommand{\ol}{\overline}

\newcommand{\ba}{\begin{eqnarray}}
\newcommand{\ea}{\end{eqnarray}}
\newcommand{\nn}{\nonumber}

\begin{document}

\begin{titlepage}

\begin{center}

\hfill UT-08-16
\vskip .5in

\textbf{\LARGE M5-brane in three-form flux\\[1ex]and multiple M2-branes}

\vskip .5in
{\large
Pei-Ming Ho$^\dagger$\footnote{
e-mail address: pmho@phys.ntu.edu.tw}, 
Yosuke Imamura$^\ddagger$\footnote{
e-mail address: imamura@hep-th.phys.s.u-tokyo.ac.jp},
Yutaka Matsuo$^\ddagger$\footnote{
e-mail address: matsuo@phys.s.u-tokyo.ac.jp}, \\
Shotaro Shiba$^\ddagger$\footnote{
e-mail address: shiba@hep-th.phys.s.u-tokyo.ac.jp}
 }\\
\vskip 3mm
{\it\large
$^\dagger$
Department of Physics and Center for Theoretical Sciences, \\
National Taiwan University, Taipei 10617, Taiwan,
R.O.C.}\\
\vskip 3mm
{\it\large
$^\ddagger$
Department of Physics, Faculty of Science, University of Tokyo,\\
Hongo 7-3-1, Bunkyo-ku, Tokyo 113-0033, Japan\\
\noindent{ \smallskip }\\
}
\vspace{60pt}
%\maketitle
\end{center}
\begin{abstract}
We investigate the Bagger-Lambert-Gustavsson model
associated with the Nambu-Poisson algebra
as a theory describing a single M5-brane.
We argue that the model is a gauge theory
associated with the volume-preserving diffeomorphism
in the three-dimensional internal space.
We derive gauge transformations,
actions, supersymmetry transformations,
and equations of motions in terms of
six-dimensional fields.
The equations of motions are written
in gauge-covariant form,
and the equations for tensor fields
have manifest self-dual structure.
We demonstrate that the double dimensional reduction
of the model reproduces the non-commutative
$U(1)$ gauge theory on a D4-brane with a small
non-commutativity parameter.
We establish relations between parameters in the BLG model
and those in M-theory.
This shows that the model describes an M5-brane
in a large $C$-field background.
% pmh
%We also argue an analogy between the model 
%and the Kodaira-Spencer theory.
%We also comment that 
%the M5-brane theory can also be understood 
%as a dynamical theory of the Nambu-Poisson structure. 

\end{abstract}

%\pacs{PACS numbers: 11.25.-w, 11.25.Mj, 11.25.Sq}%]
\end{titlepage}
%\begin{narrowtext}
\setcounter{footnote}{0}

\section{Introduction}

Recently, a model of the M5-brane world-volume field theory 
was constructed \cite{Ho:2008nn} as a system of 
infinitely many M2-branes. 
The theory of Bagger, Lambert \cite{BaggerLambert} and 
Gustavsson \cite{Gustavsson}  
was used to describe the multiple M2-brane system. 
In the BLG model, 
a background configuration of the M2-brane system 
corresponds to the choice of a Lie 3-algebra \cite{Filippov}, 
and the Lie 3-algebra used for the M5-brane is 
the Nambu-Poisson algebra \cite{Nambu} on 
a 3-manifold ${\cal N}$ which appears 
as the internal space from the M2-branes' point of view, 
but it constitutes the M5-brane world-volume 
together with the M2-brane world-volume.

It was shown \cite{Ho:2008nn} that at the quadratic order 
of the Lagrangian, 
the M5-brane theory contains a self-dual two-form gauge field, 
in addition to the scalars corresponding to 
fluctuations of the M5-brane in the transverse directions, 
as well as their fermionic super-partners. 
To the order that was computed, 
this M5-brane is different from, 
but compatible with previous formulations 
of the M5-brane theory \cite{rM51,rM52}. 
In \cite{Ho:2008nn} , higher order terms of the Lagrangian 
were not considered, 
and a truncation was applied as a short-cut to 
show the desired properties of the M5-brane model. 
In this paper, 
we show that actually the ``truncation'' did not 
really remove any physical degrees of freedom. 
The only physical components of the gauge field 
are exactly those surviving the truncation. 

By the inclusion of the nonlinear terms, 
the geometrical structure of the system 
becomes transparent.
We show that the gauge transformation defined 
by the Lie 3-algebra can be identified as the
diffeomorphism of $\mathcal{N}$ which preserve
its volume 3-form.
The gauge potential associated with
this symmetry can be identified with two-form
gauge field $b_{\mu\dot\nu}$ (an index $\mu$ for the world-volume
and another $\dot\nu$ for the internal space $\mathcal{N}$)
which is a particular combination
of the Bagger-Lambert gauge field $A_{\mu ab}$.
We show that only a particular combination of $A_{\mu ab}$
is relevant to define the gauge symmetry, the action
and the supersymmetry.  We note that the internal space
$\mathcal{N}$ may be regarded as the fiber on the three
dimensional membrane world-volume $\mathcal{M}$
in a sense.

The second characteristic feature of the system
is that not only the covariant derivative defined by
the gauge potential $b_{\mu\dot\nu}$ is covariant, 
the triplet commutator $\{X^{\dot\mu}, X^{\dot\nu},\Phi\}$
is also covariant. This follows from the fundamental identity
of the Nambu-Poisson structure.  From this combination,
we obtain the second two-form field $b_{\dot\mu\dot\nu}$
by which we can define the covariant derivative
in the fiber direction $\mathcal{N}$.
By combining two covariant derivatives, one obtains
various six dimensional field strengths associated with 
$b_{\mu\dot\nu}, b_{\dot\mu\dot\nu}$.

The BLG action and the equations of motion
are rewritten in terms of these fields.
The equations of motion for the tensor field
are written in a manifestly gauge-covariant form and
combined with the Bianchi identity
into a self-dual form.

We organize the paper as follows.
In section \ref{sec:gauge}, we derive the
BLG gauge symmetry associated with
the Nambu-Poisson bracket and identify the
gauge fields $b_{\mu\dot\nu}, b_{\dot\mu\dot\nu}$
from the gauge field $A_{\mu a b}$ and 
the scalar field $X^{\dot\mu}$. 
%the triplet commutator.  
In particular we identify the gauge transformation
as the volume-preserving diffeomorphism of $\mathcal{N}$.
In section \ref{action} we derive two types of covariant
derivatives from two-form gauge fields and 
the corresponding field strengths. 
The BLG action is then rewritten
in terms of these fields and we derive the equation of
motion.  As mentioned, it is identical to the equation
for self-dual field strength with the source terms
associated with other fields. In section \ref{SUSY},
we derive the supersymmetry transformation of
the six dimensional fields.

In \cite{Ho:2008ei}, the connection with M5 brane was used
to provide the geometrical origin of extra generators
in the construction of Lie 3-algebra which contains arbitrary
Lie algebra. In section \ref{M5D4} we provide a detailed explanation
of the derivation of D4 action from BLG model
by the double dimensional reduction.
Here the volume-preserving diffeomorphism is replaced by
the area-preserving diffeomorphism.
%After integrating out an auxiliary field,
%the action becomes Lorentz invariant.  This is not the
%feature of M5 and we believe that it provides
%a good support for our construction.
By comparing the obtained D4-brane action to the known
result,
we find explicit relations between parameters in the BLG model
and those in M-theory
(the Planck scale and the magnitude of the background $C$-field.)
This clearly indicates that the BLG model well describes
an M5-brane in a large $C$-field background.

In section \ref{geometry}, we give a few conjectural
arguments which may be helpful to understand
the geometrical nature of M5 brane in the future.
First, in section \ref{geometry} we point out that
the M5-brane theory we obtained may be 
interpreted as a dynamical theory for
the Nambu-Poisson structure. 
In this sense it is analogous to 
the Kodaira-Spencer theory \cite{BCOV} 
for the complex structure of a Calabi-Yau manifold. 

The last section is devoted to additional remarks and speculations.

For other recent developments of the BLG model, see \cite{recent}.

\section{Review of BLG model}

\paragraph{Lie 3-algebra}
The novelty of the BLG model is that it integrates a 
novel symmetry defined by Lie 3-algebra with supersymmetry.
The Lie 3-algebra is defined by an antisymmetric 
trilinear product, called Nambu bracket, 
which will be represented
by the bracket $\{*,*,*\}$.
We denote the basis of the algebra be $T^a$.
The consistency condition of Lie 3-algebra is that
it must satisfy the so-called fundamental identity:
\ba
&&\{T^a,T^b,\{T^c,T^d,T^e\}\}
=\{\{T^a,T^b,T^c\},T^d,T^e\}\nn\\
&&~~~~+\{T^c,\{T^a,T^b,T^d\},T^e\}
+\{T^c,T^d,\{T^a,T^b,T^e\}\}.
\ea
It is often convenient to define the structure constant $f^{abc}{}_d$ by
\begin{equation}
\{T^a,T^b,T^c\}=f^{abc}{}_dT^d.
\end{equation}
For the construction of an action
we need an invariant metric
\begin{equation}
h^{ab}=\langle T^a,T^b\rangle
\label{metric}
\end{equation}
which satisfies,
\ba\label{invmt}
\langle \{ T^a, T^b, T^c\}, T^d\rangle
+\langle T^c,\{T^a, T^b, T^d\}\rangle=0\,.
\ea
With the Lie 3-algebra,
various fields in BLG model
which are symbolically written as $\phi = \sum_a\phi_a T^a$
transform infinitesimally as 
\begin{equation}
\delta_{\Lambda} \phi=\sum_{a,b}\Lambda_{ab}\{T^a, T^b, \phi\},\quad
\mbox{or}\quad
\delta_\Lambda \phi_a=
\Lambda_{cd}f^{cdb}{}_a\phi_b 
\label{delf}
\end{equation}
for the gauge parameter $\Lambda_{ab}$.
The fundamental identity implies that this transformation closes
in the following sense,
\ba
[\delta_{\Lambda_1},\delta_{\Lambda_2}]\phi=
\delta_{[\Lambda_1,\Lambda_2]}\phi,\quad
[\Lambda_1,\Lambda_2]_{ab}:={\Lambda_1}_{de}{\Lambda_2}_{cb}{f^{dec}}_a+
{\Lambda_1}_{de}{\Lambda_2}_{ac}{f^{dec}}_b\,.
\ea
As a result of (\ref{invmt}), 
the metric (\ref{metric}) must also be invariant under the symmetry
(\ref{delf}),
\ba
\langle \delta_\Lambda \phi_1,\phi_2\rangle
+ \langle \phi_1, \delta_\Lambda \phi_2\rangle=0\,.
\ea
The BLG model,
whose action is constructed
with the structure constant and the invariant metric,
is a gauge theory associated with this symmetry.

\paragraph{Bagger-Lambert action}
With our notation, the action of the BLG model is given by
\begin{equation}
S=S_X+S_\Psi+S_{\rm CS}+S_{\rm int}+S_{\rm pot},
\end{equation}
where
\begin{eqnarray}
S_X&=&-\frac{1}{2}\int_{\mathcal{M}} d^3x\langle D_\mu X^I, D^\mu X^I\rangle,\\
S_{\rm pot}
&=&-\frac{1}{12}\int_{\mathcal{M}} d^3x\langle\{X^I,X^J,X^K\},\{X^I,X^J,X^K\}\rangle,\\
S_{\rm CS}
&=&\int_{\mathcal{M}} d^3x
\epsilon^{\mu\nu\lambda}\left(\frac{1}{2}
f^{abcd}A_{\mu ab}\partial_\nu A_{\lambda cd}
+\frac{1}{3}f^{cda}{}_g f^{efgb}A_{\mu ab}A_{\nu cd}A_{\lambda ef}
\right),\label{cs0}\\
S_\Psi
&=&\frac{i}{2}\int_{\mathcal{M}} d^3x\langle\ol\Psi,\Gamma^\mu D_\mu\Psi\rangle,\\
S_{\rm int}
&=&\frac{i}{4}\int_{\mathcal{M}} d^3x\langle\ol\Psi,\Gamma_{IJ}\{X^I,X^J,\Psi\}\rangle,
\end{eqnarray}
where the covariant derivatives are 
\begin{equation}
D_\mu X^I_a=\partial_\mu X^I_a-f^{bcd}{}_aA_{\mu bc}X^I_d,\quad
D_\mu\Psi_a=\partial_\mu \Psi^I_a-f^{bcd}{}_aA_{\mu bc}\Psi_d.
\label{covdex}
\end{equation}
We denote the world-volume of the membrane as $\mathcal{M}$
and its coordinate as $x^\mu$ ($\mu=0,1,2$).
The supersymmetry transformation parameter $\epsilon$ and the fermion $\Psi$
belong to ${\bf 8}_s$ and ${\bf 8}_c$ representations, respectively, of
the $SO(8)$ R-symmetry, and are represented as $32$ component spinors
satisfying
\begin{equation}
\Gamma^{\mu\nu\rho}\epsilon=+\epsilon^{\mu\nu\rho}\epsilon,\quad
\Gamma^{\mu\nu\rho}\psi=-\epsilon^{\mu\nu\rho}\psi.
\label{chiral3}
\end{equation}
This Lagrangian has a gauge symmetry associated with the 3-algebra,
\begin{equation}
\delta_{\Lambda} X^I_a= f^{bcd}{}_a\Lambda_{bc}X^I_d,\quad
\delta_{\Lambda} \Psi_a=f^{bcd}{}_a\Lambda_{bc}\Psi_d,\quad
\delta_{\Lambda} A_{\mu ab}
=D_\mu\Lambda_{ab},
\label{delbl}
\end{equation}
where
\begin{equation}
D_\mu \Lambda_{ab}
=\partial_\mu \Lambda_{ab}
- f^{cde}{}_aA_{\mu cd}\Lambda_{eb}
+f^{cde}{}_bA_{\mu cd}\Lambda_{ea}.
\label{covdell}
\end{equation}

The Bagger-Lambert action has the maximal (${\cal N}=8$) SUSY
in $d=3$,
\begin{eqnarray}
\label{eq:transX}
\delta X^I &=& i\bar\epsilon \Gamma^I \Psi, \\
\delta \Psi &=& D_\mu X^I\Gamma^\mu \Gamma^I \epsilon
-\frac{1}6 \{X^I,X^J,X^K\}\Gamma^{IJK}\epsilon, \\
\label{eq:transA}
\delta {\tilde A}_\mu{}^b{}_a &=&
i\bar{\epsilon}\Gamma_{\mu}\Gamma_I X^I_c \Psi_d f^{cdb}{}_a,\quad
\delta {\tilde A}_\mu{}^b{}_a:=f^{cdb}{}_a A_{\mu cd}.
\end{eqnarray}

%%%%%%%%%%%%%%%%%%%%%%%%%%%%%%%%%%%%%%%%%%%%%%%%%%%%%%%%%%%%%%%%%%
\section{Nambu-Poisson bracket and promotion of 3d fields to 6d}
\label{3to6}
\paragraph{Nambu-Poisson bracket as Lie 3-algebra}
For the construction of M5-brane,
we introduce an ``internal" three-manifold $\mathcal{N}$
and use the Nambu-Poisson bracket 
\ba
\left\{f,g,h\right\}_{\rm NP}=\sum_{\dot\mu\dot\nu\dot\lambda}
P^{\dot\mu\dot\nu\dot\lambda}(y)\partial_{\dot\mu}
f \partial_{\dot\nu} g \partial_{\dot\lambda} h
\ea 
on $\mathcal{N}$ as a realization of three-algebra.
Here $y^{\dot\mu}$ ($\dot\mu=\dot 1,\dot2,\dot3$) 
is the local coordinate on $\mathcal{N}$.
For literatures on the Nambu-Poisson bracket, see for example
\cite{naryLie2,Jacobian}.  One of the most important
properties of the Nambu-Poisson bracket 
is that it satisfy the analog of the fundamental identity
for arbitrary functions $f_i$ ($i=1,5$)
on $\mathcal{N}$,
\ba
&&\{f_1,f_2,\{f_3,f_4, f_5 \}_{\rm NP}\}_{\rm NP}=
\{\{f_1,f_2,f_3\}_{\rm NP},f_4,f_5\}_{\rm NP}\nn\\
&&~~~~+\{f_3,\{f_1,f_2,f_4\}_{\rm NP},f_5\}_{\rm NP}
+\{f_3,f_4,\{f_1,f_2,f_3\}_{\rm NP}\}_{\rm NP}.
\label{fi2}
\ea
This gives a very severe constraint on 
the coefficient $P^{\dot\mu\dot\nu\dot\lambda}(y)$.
Actually it is known that by the suitable choice of
the local coordinates, it can be reduced to the Jacobian,
\begin{equation} \label{NPB}
\{f,g,h\}_{\rm NP}=\epsilon^{\dot\mu\dot\nu\dot\rho}
\frac{\partial f}{\partial y^{\dot\mu}}
\frac{\partial g}{\partial y^{\dot\nu}}
\frac{\partial h}{\partial y^{\dot\rho}}.
\end{equation}
This property is referred to as the ``decomposability"
in the literature \cite{Jacobian}.
By using this fact, we can use (\ref{NPB}) 
in the following without losing generality.
We also note that the dimension of
the internal manifold $\mathcal{N}$
is essentially restricted to 3 because of the decomposability.
If we choose the basis of functions on $\mathcal{N}$ as
$\chi^a(y)$ ($a=1,2,3,\cdots$) and write the Nambu-Poisson
bracket as a Lie 3-algebra,
\ba
\{ \chi^a, \chi^b, \chi^c\}_{\rm NP}=\sum_d {f^{abc}}_d \chi^d\, ,
\ea
eq.(\ref{fi2}) implies that the structure constant 
${f^{abc}}_d$ here satisfies the fundamental identity.

The integration over the $y$-space
can be used to define the invariant metric,
\begin{equation}
\langle f,g\rangle=\frac{1}{g^2}\int_{\mathcal{N}} d^3y f(y)g(y).
\end{equation}
It is obvious that this satisfies (\ref{invmt}).
We define
\ba
h^{ab}=\langle \chi^a,\chi^b\rangle,\quad
h_{ab}=(h^{-1})_{ab}\,.
\label{metrich}
\ea
Because we have already fixed the scale of $y^{\dot\mu}$
at (\ref{NPB}), we cannot in general remove the coefficient $g$
from the metric (\ref{metrich}).
As we will show later, however,
if the internal space is ${\cal N}={\bf R}^3$,
it is possible to set this coupling at an arbitrary value
by an appropreate re-scaling of variables.

Except for the trivial case ($\mathcal{N}=\mathbf{R}^3$), 
we have to cover $\mathcal{N}$ by local patches
and the coordinates $y^{\dot\mu}$ are the local coordinates
on each patch.
If we need to go to the different patch where the local coordinates
are $y'$, the coordinate transformation between $y$ and $y'$
(say $y'^{\dot \mu}= f^{\dot\mu}(y)$)
should keep the Nambu-Poisson bracket (\ref{NPB}).
It implies that 
\ba
\{f^{\dot1}, f^{\dot2},f^{\dot3}\}=1\,.
\ea
Namely $f^{\dot\mu}(y)$ should be the volume-preserving diffeomorphism.
As we will see, the gauge symmetry of the BLG model for this
choice of Lie 3-algebra is the volume-preserving diffeomorphism 
of $\mathcal{N}$ which is very natural in this set-up.

We note that we do not need the
metric in $y^{\dot\mu}$ space.
For the definition of the theory
we only need to specify a volume form
in $\mathcal{N}$.
%The compatibility with the Nambu-Poisson bracket
%requires the volume form in the $y$-space
%to be $dy^{\dot1}\wedge dy^{\dot2}\wedge dy^{\dot3}$
%up to a constant factor.
The gauge symmetry associated with the volume-preserving
diffeomorphism
is kept not by the metric but the
various components of the self-dual two-form field
which comes out from $A_{\mu ab}$ and $X^{\dot\mu}$
(longitudinal components of $X$) as we will see.

\paragraph{Definition of 6 dim fields}
By combining the basis of $C(\mathcal{N})$,
we can treat $X_a^I(x)$ and $\Psi_a(x)$ as six-dimensional local fields
\begin{equation}
X^I(x,y)=\sum_a X^I_a(x)\chi^a(y),\quad
 \Psi(x,y)=\sum_a\Psi_a(x)\chi^a(y).
\end{equation}
Similarly, the gauge field $A_\lambda^{ab}$
can be regarded as a bi-local field:
\begin{equation}
A_\lambda(x,y,y')=A_\lambda^{ab}(x)\chi^a(y)\chi^b(y').
\end{equation}
The existence of such a bi-local field
does not mean the theory is non-local.
Let us expand it with respect
to $\Delta y^{\dot\mu}\equiv y'^{\dot\mu}-y^{\dot\mu}$ as
\begin{equation}
A_\lambda(x,y,y')
=a_\lambda(x,y)
+b_{\lambda\dot\mu}(x,y)\Delta y^{\dot\mu}
+\frac{1}{2}c_{\lambda\dot\mu\dot\nu}(x,y)\Delta y^{\dot\mu}\Delta 
y^{\dot\nu}+\cdots\,.
\label{expA}
\end{equation}
Because $A_{\lambda ab}$ always
appears in the action in the form
$f^{bcd}{}_aA_{\lambda bc}$,
the field $A_\lambda(y,y')$ is highly redundant,
and
only the component 
\begin{equation}\label{blmud}
b_{\lambda\dot\mu}(x,y)
=\left.\frac{\partial}{\partial y'^{\dot\mu}}A_\lambda(x,y,y')
\right|_{y'=y}
\end{equation}
contributes to the action\footnote{
In \cite{Ho:2008nn} it was treated as a trick 
(or an approximation by neglecting the irrelevant parts)
to derive M5 action.  However, it turned out that this
is actually the exact statement.}.
For example, the covariant derivative (\ref{covdex}) of BLG model
is rewritten for our case as,
\begin{eqnarray}
D_\lambda X^I(x,y)
&\equiv&(\partial_\lambda X^{Ia}(x)-gf^{bcd}{}_aA_{\lambda bc}X^I_d(x))\chi^a(y)
\nonumber\\
&=&
\partial_\lambda X^I(x,y)
-g\epsilon^{\dot\mu\dot\nu\dot\rho}
\left.
\frac{\partial^2 A_\lambda(x,y,y')}{\partial y^{\dot\mu}\partial y'^{\dot\nu}}
\right|_{y=y'}\frac{\partial X^I(x,y)}{\partial y^{\dot\rho}}
\nonumber\\
&=&
\partial_\lambda X^I(x,y)
-g\epsilon^{\dot\mu\dot\nu\dot\rho}
(\partial_{\dot\mu}b_{\lambda\dot\nu}(x,y))
(\partial_{\dot\rho}X^I(x,y))\nn\\
&=& \partial_\lambda X^I
-g\{b_{\lambda\dot\nu},y^{\dot\nu},X^I\}.
\label{covdel}
\end{eqnarray}
%Between the first and second lines in
%(\ref{covdel}), 
%we used the definition of the Nambu-Poisson bracket
%\begin{equation}
%f^{abc}{}_d g_ah_bf_c\chi^d(y)
%=\{g(y),h(y),f(y)\}
%=\epsilon^{\dot\mu\dot\nu\dot\rho}
%\left.\frac{\partial g(y)}{\partial y^{\dot\mu}}
%\frac{\partial h(y')}{\partial y'^{\dot\nu}}
%\frac{\partial f(y'')}{\partial y''^{\dot\rho}}
%\right|_{y=y'=y''}
%\end{equation}
%with the product of two functions
%$g(y)h(y')$ replaced by $A_\lambda(y,y')$.
The covariant derivative for the fermion field is similarly,
\ba
D_\lambda \Psi(x,y)=\partial_\lambda \Psi(x,y) 
-g\epsilon^{\dot\mu\dot\nu\dot\rho}
(\partial_{\dot\mu}b_{\lambda\dot\nu}(x,y))
(\partial_{\dot\rho}\Psi(x,y))=
\partial_\lambda \Psi
-g\{b_{\lambda\dot\nu},y^{\dot\nu},\Psi\}\,.
\ea

\paragraph{Longitudinal fields}
In \cite{Ho:2008nn}, this theory
written in terms of  fields
on six dimensions is identified with the
theory describing a single M5-brane.
At this point,
only the $x^\mu$ part of the metric $g_{\mu\nu}=\eta_{\mu\nu}$
is defined,
and we still have $SO(8)$ global symmetry,
which is different from the $SO(5)$ symmetry
expected in the M5-brane theory.

This is quite similar to the situation
in which we consider the D-brane Born-Infeld action.
The Born-Infeld D$p$-brane action
of ten-dimensional superstring theory
possesses $SO(1,9)$ Lorentz symmetry regardress of the
world-volume dimension $p+1$.
The rotational symmetry is reduced to $SO(9-p)$
for the transverse directions only after
fixing the general coordinate transformation symmetry on the
world-volume with the static gauge condition
\footnote{
Turning on a background field such as the $B$-field 
will of course also break the global symmetry. 
For the discussion here we are treating the background fields 
as covariant dynamical fields. 
}
\begin{equation}
X^\mu(\sigma)=\sigma^\mu.
\label{staticgauge}
\end{equation}
This gauge fixing breaks the global symmetry from $SO(1,9)$
to $SO(9-p)$,
and at the same time the world-volume metric
is induced from the target space metric
through (\ref{staticgauge}).

We can interpret the six-dimensional theory
we are considering here as
a theory obtained from an $SO(1,10)$ symmetric
covariant theory by taking a partial
static gauge for three among six world-volume coordinates.
As we mentioned above, however,
we do not have full diffeomorphism in the $y^{\dot\mu}$ space.
The action is invariant only under volume-preserving diffeomorphism.
This implies that we cannot completely fix the fields $X^{\dot\mu}$,
and there are remaining physical degrees of freedom.
For this reason, we should loosen the static gauge condition
as \cite{Ho:2008nn}
\begin{equation}
X^{\dot\mu}(x,y)=y^{\dot\mu}
+b^{\dot\mu}(x,y),\quad
b_{\dot\mu\dot\nu}
=\epsilon_{\dot\mu\dot\nu\dot\rho}b^{\dot\rho}.
\label{xmuans}
\end{equation}
As was shown in \cite{Ho:2008nn},
the tensor field $b_{\dot\mu\dot\nu}$
is identified with
a part of the $2$-form gauge field on a M5-brane.

%%%%%%%%%%%%%%%%%%%%%%
\paragraph{Comments on the coupling constant}
In the case of ordinary Yang-Mills theories,
there are two widely-used conventions for
coupling constants and normalization of gauge fields.
One way is to normalize a gauge field by
the canonical kinetic term $-(1/4)F_{\mu\nu}^2$ and put the
coupling constant in the covariant derivative $D=d-igA$.
The other choice is to define the covariant derivative
$D=d-iA$ without using the coupling constant and instead
put $1/g^2$ in front of the kinetic term of the gauge field.
Similarly, there are different conventions
for coupling constant in the case of the BL theory, too.
In the above, we put the coupling constant $g$ in
the definition of the metric (\ref{metrich}).
This corresponds to the second convention we mentioned above.
We can move the coupling dependence from the
overall factor to the interaction terms
by re-scaling the fields
\begin{equation}
X^I\rightarrow c X^I,\quad
\Psi\rightarrow c\Psi,\quad
b_{\mu\dot\mu}\rightarrow cb_{\mu\dot\mu},
\label{rescale1}
\end{equation}
with $c=g$.
In general, as ordinary Yang-Mills theories,
we cannot remove the coupling constant completely from
the action.

If the internal space ${\cal N}$ is ${\bf R}^3$, however,
we have an extra degree of freedom for re-scaling,
and it is in fact possible to the coupling constant from the action.
Let us consider the following re-scaling of variabvles.
\begin{equation}
X^I\rightarrow c'^3X^I,\quad
\Psi\rightarrow c'^3\Psi,\quad
b_{\mu\dot\mu}\rightarrow c'^4b_{\mu\dot\mu},\quad
y^{\dot\mu}\rightarrow c'^2y^{\dot\mu}.
\label{rescale2}
\end{equation}
This variable change is associated with an outer automorphism
of the algebra, and does not change the relative coefficients
in the action.
The only change in the action is the overall factor.
We can thus absorb the coupling constant by (\ref{rescale2}),
and this implies that the six-dimensional theory does not have
any coupling constant.

We can adopt an elegant convention in which no coupling constant
appears.
However, we adopt a different convention below.
Because we interpret the six-dimensional theory
as a theory of an M5-brane,
we would like to regard the scalar field $X^I$
as the coordinates of the target space with
mass dimension $-1$.
We also give the meaning to the variables $y^{\dot\mu}$ as
the world-volume coordinates, which also have mass dimension
$-1$.
We choose the parametrization in
the $y^{\dot\mu}$ space so that the
linear part of the six-dimensional action is invariant
under Lorentz transformations in the $(x^\mu,y^{\dot\mu})$ space.
After fixing the scale of $X^I$ and $y^{\dot\mu}$ in this way,
we can no longer use the two re-scalings (\ref{rescale1}) and (\ref{rescale2})
to change the coupling constant
and overall coefficient of the action.
These two parameters have physical meaning now.

In the following, in order to express the coupling constant dependence
of each term in the action clearly,
we separate the coupling constant $g$ from the
structure constant.
We also introduce an overall coefficient $T_6$,
which is regarded as an effective tension
of the M5-brane.
This plays an important role in the parameter matching
in \S\ref{M5D4}, but
we will omit this factor
in \S\ref{sec:gauge}, \S\ref{action},
and \S\ref{SUSY} for simplicity
because it is irrelevant to the analysis
in these sections.

%%%%%%%%%%%%%%%%%%%%%%%%%%%%%%%%%%%%%%%%%%%%%%%%%%%%%%%%%%
\section{Gauge symmetry of M5 from Lie 3-algebra}
\label{sec:gauge}

\paragraph{Gauge transformaion}
The gauge transformations of the scalar fields $X^I$
and fermion fields $\Psi$ are given by
\ba
&& \delta_{\Lambda}X^I(x,y)
=g\Lambda_{ab}(x){f^{abc}}_d X^I_c(x) \chi^d(y)\nn\\
&&~~~~=
g\Lambda_{ab}(x)\{\chi^a, \chi^b, X^I\}=
g(\delta_\Lambda y^{\dot\rho})
\partial_{\dot\rho}X^I(x,y),\nn\\
&&\delta_{\Lambda}\Psi(x,y)
=g\Lambda_{ab}(x)\{\chi^a, \chi^b, \Psi\}
= g(\delta_\Lambda y^{\dot\rho})
\partial_{\dot\rho}\Psi(x,y),
\label{gaugetr}
\ea
where we used
\ba
{f^{abc}}_d=\langle \{\chi^a,\chi^b,\chi^c\}, \chi_d\rangle\,,\quad
\sum_{a}\chi^a(y) \chi_a(y')=\delta^{(3)}(y-y')\,.
\ea
$\delta_\Lambda y^{\dot\mu}$ is defined as
\ba
&& \delta_{\Lambda} y^{\dot\lambda}=
\epsilon^{\dot\lambda\dot\mu\dot\nu}\partial_{\dot\mu}
\Lambda_{\dot\nu}(x,y),\\
&& \Lambda_{\dot\mu}(x,y)=\partial'_{\dot\mu}\tilde\Lambda(x,y,y')|_{y'=y}\,,\quad
\tilde\Lambda(x,y,y'):=\Lambda_{ab}(x)\chi^a(y)\chi^b(y').
\ea
We note that
although the parameter of a gauge transformation
may be expressed as a bi-local function
$\tilde\Lambda(x,y,y')$,
the gauge transformation induced by it
depends only on its component $\Lambda_{\dot\mu}(x,y)$
which is local in $\mathcal{N}$.
It comes from the fact that the gauge transformation
by $\Lambda_{ab}$  is always defined through
the combination ${f^{abc}}_d \Lambda_{ab}$\,.

The same argument can be applied to the gauge field
$A_{\mu}(x,y,y')$.  As we already mentioned, since it appears
only through the combination $A_{\mu a b}{f^{abc}}_d$, 
the local field $b_{\mu\dot\lambda}(x,y)$ defined as (\ref{blmud})
shows up in the action.

The transformation (\ref{gaugetr}) may be regarded as
the infinitesimal reprametrization
\begin{equation}
y'^{\dot\lambda}=y^{\dot\lambda}
-g \delta y^{\dot\lambda}.
\label{coordtr}
\end{equation}
Since $\partial_{\dot\mu} \delta y^{\dot\mu}=0$,
it represents the volume-preserving diffeomorphism.
Since the symmetry is local on $\mathcal{M}$, the gauge
parameter is an arbitrary function of $x$.
So what we have obtained is a gauge theory on $\mathcal{M}$
whose gauge group is the volume-preserving 
diffeomorphism of $\mathcal{N}$.
In this sense, the world-volume of M5 brane may be regarded
as the vector bundle $\mathcal{N}\rightarrow \mathcal{M}$
but the gauge transformation on each fiber is not merely the
linear transformation but the diffeomorphism 
on the fiber which preserves
the volume form 
\begin{equation}
\omega
=dy^{\dot 1}
\wedge dy^{\dot 2}
\wedge dy^{\dot 3}\,.
\label{vform}
\end{equation}

As we mentioned in the previous section,
among eight scalar fields $X^I$,
the last five components $X^i$
are treated as scalar fields
representing the transverse fluctuations of
the M5-brane.
The other three $X^{\dot\mu}$ (longitudinal field)
are rewritten as as
\begin{equation}
X^{\dot\mu}(y)
=\frac{y^{\dot\mu}}{g}
+
\frac{1}{2}\epsilon^{\dot\mu\dot\kappa\dot\lambda}
b_{\dot\kappa\dot\lambda}(y).
\label{xeqyb}
\end{equation}
We chose the coefficients
so that we obtain
Lorentz invariant
kinetic terms in the six-dimensional action.
The gauge transformation of $b_{\dot\mu\dot\nu}$ can be
derived from
(\ref{gaugetr})
and (\ref{xeqyb}) as
\begin{equation}
\delta_{\Lambda}b_{\dot\kappa\dot\lambda}(y)
=\partial_{\dot\kappa}\Lambda_{\dot\lambda}
-\partial_{\dot\lambda}\Lambda_{\dot\kappa}
+g(\delta_{\Lambda} y^{\dot\rho})
\partial_{\dot\rho} b_{\dot\kappa\dot\lambda}(y) .
\end{equation}

The gauge transformation of the gauge field $A_\lambda(x,y,y')$
is given by
$\delta_{\Lambda} A_\lambda(x,y,y')
=D_\lambda\tilde\Lambda(x, y,y')$.
The covariant derivative of a bi-local field
is defined by tensoring the covariant derivative (\ref{covdel}) for a local field,
and we obtain
\begin{equation}
D_\lambda\Lambda(y,y')
=
\partial_\lambda\Lambda(y,y')
-g\epsilon^{\dot\mu\dot\nu\dot\rho}
[\partial_{\dot\mu}b_{\lambda\dot\nu}(y)
\partial_{\dot\rho}\Lambda(y,y')
+\partial'_{\dot\mu}b_{\lambda\dot\nu}(y')
\partial'_{\dot\rho}\Lambda(y,y')].
\label{ayygaugeyttr}
\end{equation}
From this we can extract the
transformation law of the
component field
$b_{\lambda\dot\sigma}$
\begin{equation}
\delta_{\Lambda} b_{\lambda\dot\sigma}
=
\partial'_{\dot\mu}\delta_{\Lambda} A_\lambda(y,y')|_{y'=y}
=\partial_\lambda\Lambda_{\dot\sigma}
-g\partial_{\dot\sigma}\xi_\Lambda
-g\delta_{\rm gc}b_{\lambda\dot\sigma},
\label{xitr}
\end{equation}
where $\delta_{\rm gc}b_{\lambda\dot\sigma}$ is the
coordinate transformation in $y$-space
\begin{equation}
\delta_{\rm gc}b_{\lambda\dot\sigma}
=-\delta_\Lambda y^{\dot\tau}\partial_{\dot\tau}b_{\lambda\dot\sigma}
-(\partial_{\dot\sigma}\delta_\Lambda y^{\dot\tau})b_{\lambda\dot\tau},
\label{gcb}
\end{equation}
and
$\xi_\Lambda$ is defined by
\begin{equation}
\xi_\Lambda
=
\epsilon^{\dot\mu\dot\nu\dot\rho}
(\partial_{\dot\mu}b_{\lambda\dot\nu}\Lambda_{\dot\rho}
+b_{\lambda\dot\mu}\partial_{\dot\nu}\Lambda_{\dot\rho}).
\end{equation}

In addition to these gauge transformations derived from
(\ref{covdex}) and (\ref{covdell}),
there is an additional gauge transformation which acts
only on the field $b_{\lambda\dot\mu}$.
As we can see in (\ref{covdel}),
$b_{\lambda\dot\mu}$ appears in the covariant derivative
in the form of the rotation in the $y^{\dot\mu}$ space.
This means that $D_\mu\Phi$
is invariant under
\begin{equation}
\delta b_{\lambda\dot\mu}=-\partial_{\dot\mu}\Lambda_\lambda.
\label{gaugetr2}
\end{equation}
We can easily check that
the Chern-Simons term is also invariant under this transformation,
and thus (\ref{gaugetr2}) is also a gauge symmetry of the theory.

Now we summarize the gauge transformation of the
six-dimensional theory.
\begin{eqnarray}
\delta_{\Lambda}X^i
&=&g(\delta_\Lambda y^{\dot\rho})\partial_{\dot\rho}X^i, \label{gt1}
\\
\delta_{\Lambda}\Psi
&=&g(\delta_\Lambda y^{\dot\rho})\partial_{\dot\rho}\Psi,\\
\delta_{\Lambda}b_{\dot\kappa\dot\lambda}
&=&\partial_{\dot\kappa}\Lambda_{\dot\lambda}
-\partial_{\dot\lambda}\Lambda_{\dot\kappa}
+g(\delta_\Lambda y^{\dot\rho})\partial_{\dot\rho} b_{\dot\kappa\dot\lambda},\\
\delta_{\Lambda} b_{\lambda\dot\sigma}
&=&\partial_\lambda\Lambda_{\dot\sigma}
-\partial_{\dot\sigma}\Lambda_\lambda
-g\delta_{\rm gc}b_{\lambda\dot\sigma}. \label{gt4}
\end{eqnarray}
We absorbed $\xi_\Lambda$ in (\ref{xitr})
into the definition of the parameter $\Lambda_\lambda$.
In the weak coupling limit $g\rightarrow 0$,
we obtain the standard gauge tramnsformation on an M5-brane.

\paragraph{Covariant derivatives in 6 dim}

An intriguing feature of our six dimensional model is
that one may define the covariant derivative
in the {\em fiber} direction.

By using the fundamental identity,
it is easy to show that
if $\Phi_1$, $\Phi_2$, and $\Phi_3$ are covariant fields
(such as $X^I$ or $\Psi$), not only 
$D_\mu\Phi_1$ but $\{\Phi_1,\Phi_2,\Phi_3\}$ are also covariant
because of the fundamental identity,
\ba
\delta_\Lambda\{\Phi_1,\Phi_2,\Phi_3\}=
\{\delta_\Lambda\Phi_1,\Phi_2,\Phi_3\}+
\{\Phi_1,\delta_\Lambda\Phi_2,\Phi_3\}+
\{\Phi_1,\Phi_2,\delta_\Lambda\Phi_3\}\,.
\ea
It implies that the following combination 
defines the ``covariant" derivative along
the fiber direction,
\begin{eqnarray}
{\cal D}_{\dot\mu}\Phi
&\equiv&\frac{g^2}{2}\epsilon_{\dot\mu\dot\nu\dot\rho}
\{X^{\dot\nu},X^{\dot\rho},\Phi\}
\nonumber\\
&=&
\partial_{\dot\mu}\Phi
+g(
\partial_{\dot\lambda}b^{\dot\lambda}\partial_{\dot\mu}\Phi
-\partial_{\dot\mu}b^{\dot\lambda}\partial_{\dot\lambda}\Phi
)
+\frac{g^2}{2}\epsilon_{\dot\mu\dot\nu\dot\rho}
\{b^{\dot\nu},b^{\dot\rho},\Phi\}.\label{ddotmu}
\end{eqnarray}
Together with (\ref{covdel}), which we repeat here again,  
\ba
{\cal D}_\mu\Phi
&\equiv&D_\mu\Phi
=\partial_\mu\Phi
-g\{b_{\mu\dot\nu},y^{\dot\nu},\Phi\},\label{dmu}
\ea
we have a set of covariant derivatives on M5 world-volume.

These covariant derivatives possess the
following important properties.
\begin{itemize}
\item Leibniz rule:
\begin{equation}
{\cal D}_{\ul\mu}\{\Phi_1,\Phi_2,\Phi_3\}
=
\{{\cal D}_{\ul\mu}\Phi_1,\Phi_2,\Phi_3\}
+\{\Phi_1,{\cal D}_{\ul\mu}\Phi_2,\Phi_3\}
+\{\Phi_1,\Phi_2,{\cal D}_{\ul\mu}\Phi_3\} .
\end{equation}
\item
Integration by parts:
\begin{equation}
\int d^3xd^3y \Phi_1{\cal D}_{\ul\mu}\Phi_2
=-\int d^3xd^3y ({\cal D}_{\ul\mu}\Phi_1)\Phi_2 .
\end{equation}
\end{itemize}
Here ${\cal D}_{\ul \mu}$ ($\ul\mu=0,1,\cdots,5$)
represents ${\cal D}_\mu$
and ${\cal D}_{\dot\mu}$.

\paragraph{Field strength}

As special cases of these covariant derivatives,
we define the following field strengths of the tensor field. 
\begin{eqnarray}
{\cal H}_{\lambda\dot\mu\dot\nu}
&=&\epsilon_{\dot\mu\dot\nu\dot\lambda}{\cal D}_\lambda X^{\dot\lambda}
\nonumber\\
&=&H_{\lambda\dot\mu\dot\nu}
-g\epsilon^{\dot\sigma\dot\tau\dot\rho}
(\partial_{\dot\sigma}b_{\lambda\dot\tau})
\partial_{\dot\rho}b_{\dot\mu\dot\nu},\label{h12def}\\
{\cal H}_{\dot1\dot2\dot3}
&=&g^2\{X^{\dot1},X^{\dot2},X^{\dot3}\}-\frac{1}{g}
=\frac{1}{g}(V-1)
\nonumber\\
&=&H_{\dot1\dot2\dot3}
+\frac{g}{2}
(\partial_{\dot\mu}b^{\dot\mu}\partial_{\dot\nu}b^{\dot\nu}
-\partial_{\dot\mu}b^{\dot\nu}\partial_{\dot\nu}b^{\dot\mu})
+g^2\{b^{\dot1},b^{\dot2},b^{\dot3}\}, 
\label{h30def}
\end{eqnarray}
where $V$ is the ``induced volume''
\begin{equation}
V=g^3\{X^{\dot1},X^{\dot2},X^{\dot3}\},
\end{equation}
and $H$ is the linear part of the field strength
\begin{eqnarray}
H_{\lambda\dot\mu\dot\nu}
&=&
\partial_{\lambda}b_{\dot\mu\dot\nu}
-\partial_{\dot\mu}b_{\lambda\dot\nu}
+\partial_{\dot\nu}b_{\lambda\dot\mu},\\
H_{\dot\lambda\dot\mu\dot\nu}
&=&
\partial_{\dot\lambda}b_{\dot\mu\dot\nu}
+\partial_{\dot\mu}b_{\dot\nu\dot\lambda}
+\partial_{\dot\nu}b_{\dot\lambda\dot\mu}.
\end{eqnarray}
${\cal H}$ are covariantly transformed under the gauge transformation.

Just like the case of ordinary gauge theories,
the field strength ${\cal H}$ arises
in the commutator of the covariant derivatives
defined above: 
\begin{eqnarray}
[{\cal D}_{\dot\mu},{\cal D}_{\dot\nu}]\Phi
&=&
g^2
\epsilon_{\dot\nu\dot\mu\dot\sigma}
\{{\cal H}_{\dot1\dot2\dot3},X^{\dot\sigma},\Phi\},
\label{comm1}\\
{}[{\cal D}_\lambda,{\cal D}_{\dot\lambda}]\Phi
&=&
g^2
\{{\cal H}_{\lambda\dot\nu\dot\lambda},X^{\dot\nu},\Phi\},
\label{comm2}\\
{}[{\cal D}_\mu,{\cal D}_\nu]\Phi
&=&-\frac{g}{V}
\epsilon_{\mu\nu\lambda}{\cal D}_\rho\wt{\cal H}^{\rho\lambda\dot\kappa}
{\cal D}_{\dot\kappa}\Phi,
\label{comm3}
\end{eqnarray}
where
the dual field strength
$\wt{\cal H}$ is defined by
\begin{equation}
\wt{\cal H}^{\lambda\rho\dot\kappa}
=\frac{1}{2}\epsilon^{\lambda\rho\dot\kappa\sigma\dot\mu\dot\nu}
{\cal H}_{\sigma\dot\mu\dot\nu},
\quad
\wt{\cal H}^{\mu\nu\rho}
=\frac{1}{6}\epsilon^{\mu\nu\rho\dot\mu\dot\nu\dot\rho}{\cal H}_{\dot\mu\dot\nu\dot\rho}.
\end{equation}

\section{M5 action and equation of motion}
\label{action}

%\paragraph{(24) 6 dim action for $X$ and $\Psi$}
We rewrite the various parts of the Bagger-Lambert
action in terms of the six dimensional fields and
their covariant derivatives,
\begin{eqnarray}
S_X+S_{\rm pot}
&=&\int d^3x
\left\langle
-\frac{1}{2}({\cal D}_\mu X^i)^2
-\frac{1}{2}({\cal D}_{\dot\lambda}X^i)^2
-\frac{1}{4}{\cal H}_{\lambda\dot\mu\dot\nu}^2
-\frac{1}{12}{\cal H}_{\dot\mu\dot\nu\dot\rho}^2
\right.\nonumber\\&&\left.
-\frac{1}{2g^2}
-\frac{g^4}{4}\{X^{\dot\mu},X^i,X^j\}^2
-\frac{g^4}{12}\{X^i,X^j,X^k\}^2\right\rangle,\\
S_\Psi+S_{\rm int}
&=&\int d^3x\left\langle
\frac{i}{2}\ol\Psi\Gamma^\mu {\cal D}_\mu\Psi
+\frac{i}{2}\ol\Psi\Gamma^{\dot\rho}\Gamma_{\dot1\dot2\dot3}{\cal D}_{\dot\rho}\Psi
\right.\nonumber\\&&\left.
+\frac{ig^2}{2}\ol\Psi\Gamma_{\dot\mu i}\{X^{\dot\mu},X^i,\Psi\}
+\frac{ig^2}{4}\ol\Psi\Gamma_{ij}\{X^i,X^j,\Psi\}
\right\rangle.
\label{Sfermi0}
\end{eqnarray}
The scalar kinetic term is manifestly Lorentz symmetric
up to the different structure inside the covariant derivatives
${\cal D}_\mu$ and ${\cal D}_{\dot\mu}$.
The Chern-Simons term
cannot be rewritten in
manifestly gauge-covariant form.
\begin{eqnarray}
S_{\rm CS}
&=&
\int d^3x
\epsilon^{\mu\nu\lambda}
\left\langle
-\frac{1}{2}
\epsilon^{\dot\mu\dot\nu\dot\lambda}
\partial_{\dot\mu}b_{\mu\dot\nu}\partial_\nu b_{\lambda\dot\lambda}
\right.\nonumber\\&&\hspace{2cm}\left.
+\frac{g}{6}\epsilon^{\dot\mu\dot\nu\dot\lambda}
\partial_{\dot\mu}b_{\nu\dot\nu}
\epsilon^{\dot\rho\dot\sigma\dot\tau}
\partial_{\dot\sigma}b_{\lambda\dot\rho}
(\partial_{\dot\lambda}b_{\mu\dot\tau}-\partial_{\dot\tau}b_{\mu\dot\lambda})
\right\rangle\nonumber\\
&=&
\int d^3x\int_y
\epsilon^{\mu\nu\lambda}
\left(
-\frac{1}{2}db_\mu\wedge\partial_\nu b_\lambda
-\frac{g}{6}
(*db_\mu)
\wedge(*db_\nu)
\wedge(*db_\lambda)
\right).
\label{CSt}
\end{eqnarray}
In the second expression we treat $b_{\mu\dot\mu}$
as a one-form field $b_\mu=b_{\mu\dot\mu}dy^{\dot\mu}$
in the $y$-space.  However, the equation of motion
which is derived from these actions turns out to
be manifestly gauge-covariant.

\paragraph{Comments on fermion action}
In the fermion kinetic terms
in (\ref{Sfermi0}), only the
$SO(1,2)\times SO(3)$ subgroup of the Lorentz
symmetry is manifest due to the
existence of $\Gamma_{\dot1\dot2\dot3}$ in one of
two terms.
We can remove this unwanted factor
from the kinetic term
by the unitary transformation
\begin{equation}
\ol\Psi=\ol\Psi'U,\quad
\Psi=U\Psi',
\label{unitary}
\end{equation}
where $U$ is the matrix
\begin{equation}
U
=\exp\left(-\frac{\pi}{4}\Gamma_{\dot1\dot2\dot3}\right)
=\frac{1}{\sqrt2}(1-\Gamma_{\dot1\dot2\dot3}).
\end{equation}
The SUSY parameter $\epsilon$ is also transformed in the same way.
Note that both $\Psi$ and $\ol\Psi$ are transformed by $U$.
This is consistent with the Dirac conjugation.
As the result of the unitary transformation, the fermion terms
in the action become
\begin{eqnarray}
S_\Psi+S_{\rm int}
&=&\int d^3x\left\langle
\frac{i}{2}\ol\Psi'\Gamma^\mu {\cal D}_\mu\Psi'
+\frac{i}{2}\ol\Psi'\Gamma^{\dot\rho}{\cal D}_{\dot\rho}\Psi'
\right.\nonumber\\&&\left.
+\frac{ig^2}{2}\ol\Psi'\Gamma_{\dot\mu i}\{X^{\dot\mu},X^i,\Psi'\}
-\frac{ig^2}{4}\ol\Psi'\Gamma_{ij}\Gamma_{\dot1\dot2\dot3}\{X^i,X^j,\Psi'\}
\right\rangle.
\end{eqnarray}
%Unfortunately, we still have $\Gamma_{\dot1\dot2\dot3}$ in the
%interaction term.
After the unitary transformation,
the condirion
(\ref{chiral3})
becomes the chirality condition in six dimension, 
\begin{equation}
\Gamma^7\epsilon'=\epsilon',\quad
\Gamma^7\Psi'=-\Psi',
\end{equation}
where the chiality matrix $\Gamma^7$ is defined by
\begin{equation}
\Gamma^{\mu\nu\rho}\Gamma^{\dot1\dot2\dot3}=\epsilon^{\mu\nu\rho}\Gamma^7.
\end{equation}
This means that
the supersymmetry realized in this theory
is the chiral ${\cal N}=(2,0)$ supersymmetry,
which is the same as the supersymmetry on an M5-brane.

\paragraph{Equations of motion}
It is easy to obtain 
the equations of motion for the scalar fields and fermion field as 
\begin{eqnarray}
0&=&{\cal D}_\mu^2X^i+{\cal D}_{\dot\mu}^2X^i
\nonumber\\&&
+g^4\{X^{\dot\mu},X^j,\{X^{\dot\mu},X^j,X^i\}\}
+\frac{g^4}{2}\{X^j,X^k,\{X^j,X^k,X^i\}\}
\nonumber\\
&&+\frac{ig^2}{2}\{\ol\Psi'\Gamma_{\dot\mu i},X^{\dot\mu},\Psi'\}
+\frac{ig^2}{2}\{\ol\Psi'\Gamma_{ij}\Gamma_{\dot1\dot2\dot3},X^j,\Psi'\}
,\\
0&=&\Gamma^\mu \cD_\mu \Psi'
+\Gamma^\drho \cD_\drho\Psi'
+g^2\Gamma_{\dmu i}\{X^\dmu,X^i,\Psi'\}
-\frac{g^2}2\Gamma_{ij}\Gamma_{\dot1\dot2\dot3}\{X^i,X^j,\Psi'\}.
\label{feom}
\end{eqnarray}
The equations of motion of gauge fields $b_{\mu\dot\mu}$ and $b_{\dot\mu\dot\mu}$,
and the bianchi identity are combined
into the self-dual form: 
\begin{eqnarray}
&&{\cal D}_{\lambda}{\cal H}^{\lambda\dot\mu\dot\nu}
+{\cal D}_{\dot\lambda}{\cal H}^{\dot\lambda\dot\mu\dot\nu}
=gJ^{\dot\mu\dot\nu},\label{maxwell1}\\
&&{\cal D}_\lambda\wt{\cal H}^{\lambda\mu\dot\nu}
+{\cal D}_{\dot\lambda}{\cal H}^{\dot\lambda\mu\dot\nu}
=gJ^{\mu\dot\nu},\label{maxwell2}\\
&&{\cal D}_\lambda\wt{\cal H}^{\lambda\mu\nu}
+{\cal D}_{\dot\lambda}\wt{\cal H}^{\dot\lambda\mu\nu}
=0.\label{maxwell3}
\end{eqnarray}
The first two are equations of motion
obtained from the action, while the last one
is a Bianchi identity derived from the commutation relation
(\ref{comm3}).
The currents on the right hand sides are given by
\begin{eqnarray}
J^{\dot\rho\dot\sigma}
&=&g
(\{X^i,{\cal D}_{\dot\sigma}X^i,X^{\dot\rho}\}
-(\dot\rho\leftrightarrow\dot\sigma))
-\frac{g^3}{2}
\epsilon^{\dot\rho\dot\sigma\dot\mu}
\{X^i,X^j,\{X^i,X^j,X^{\dot\mu}\}\}
\nonumber\\&&
+\frac{ig}{2}
(\{\ol\Psi'\Gamma^{\dot\sigma},X^{\dot\rho},\Psi'\}
-(\dot\rho\leftrightarrow\dot\sigma))
+\frac{ig}{2}\epsilon^{\dot\rho\dot\sigma\dot\mu}\{\ol\Psi'\Gamma_{\dot\mu i},X^i,\Psi'\},\\
J^{\mu\dot\nu}
&=&g\{X^i,{\cal D}_\mu X^i,X^{\dot\nu}\}
+\frac{ig}{2}\{\ol\Psi'\Gamma^\mu,\Psi',X^{\dot\nu}\}.
\end{eqnarray}

The self-dual tensor field ${\cal H}$,
chiral fermion field $\Psi'$,
and the five scalar fields $X^i$ form a
tensor multiplet of ${\cal N}=(2,0)$ supersymmetry \cite{HST},
which is the same as the field contents on an M5-brane.

%%%%%%%%%%%%%%%%%%%%%%%%%%%%%%%%%%%%%%%%%%%%%%%%%%%%%%%%%%%%%%%%%%%%%%%%%%
\section{Supersymmetry}
\label{SUSY}
In this section we rewrite the supersymmetry transformations
(\ref{eq:transX})-(\ref{eq:transA}) in terms of the
six-dimensional covariant derivatives
and field strength.
The transformation law (\ref{eq:transA})
of the gauge field $A_{\mu ab}$
with coupling constant inderted is
\begin{equation}
\label{eq:transA2}
\wt A_\mu{}^b{}_a=
ig\bar{\epsilon}\Gamma_{\mu}\Gamma_I X^I_c \Psi_d f^{cdb}{}_a.
\end{equation}
We cannot determine uniquely the transformation law of
the component field
$b_{\mu\dnu}$ from this equation
because of the existence of the
gauge transformation (\ref{gaugetr2}), which acts only on $b_{\mu\dot\mu}$.
In fact, the transformation (\ref{eq:transA2}) only gives
\begin{equation}
\delta \left(\epsilon^{\dmu\dnu\drho}\partial_\dmu b_{\lambda\dnu}\partial_\drho f(y) \right)
= ig\bar\epsilon\Gamma_\lambda\Gamma_I\{X^I,\Psi,f(y)\},
\label{transdb}
\end{equation}
where $f(y)$ is an arbitrary function of $y^\dmu$.
One possible choice for $\delta b_{\mu\dot\mu}$ is
\begin{equation}
\label{eq:transb}
\delta b_{\mu\dnu}
=ig(\bar\epsilon \Gamma_I\Gamma_\mu  \Psi)\partial_\dnu X^I.
\end{equation}
We can easily check that this transformation law
reproduces (\ref{transdb}).

In some situations
%as we will mention in Sec. \ref{geometry},
an explicit appearance of $b_{\mu\dot{\mu}}$ is 
not necessary, 
but all we need is
$B_{\mu}{}^{\dot{\mu}}\equiv \epsilon^{\dot\mu\dot\nu\dot\rho}\partial_{\dot\nu}b_{\mu\dot\rho}$,
which satisfies the constraint $\partial_{\dot{\mu}} B_{\mu}{}^{\dot{\mu}} = 0$.
The SUSY transformation for $B_{\mu}{}^{\dot{\mu}}$ is 
uniquely determined from (\ref{transdb}) as
\begin{equation}
\delta B_{\mu}{}^{\dot{\mu}} =
i g\bar{\epsilon} \Gamma_{\mu}\Gamma_I 
\epsilon^{\dot{\mu}\dot{\nu}\dot{\lambda}} 
\partial_{\dot{\nu}}X^I \partial_{\dot{\lambda}} \Psi, 
\end{equation}
and it is obvious that the constraint is SUSY invariant, i.e.
\begin{equation}
\delta (\partial_{\dot{\mu}} B_{\mu}{}^{\dot{\mu}} ) = 0.
\end{equation}

The transformation laws rewritten
in terms of the six-dimensional notation are
\begin{eqnarray}
\delta X^i
&=&i\ol\epsilon'\Gamma^i\Psi',\\
\delta \Psi'
&=&{\cal D}_\mu X^i\Gamma^\mu\Gamma^i\epsilon'
+{\cal D}_{\dot\mu}X^i\Gamma^{\dot\mu}\Gamma^i\epsilon'
\nonumber\\&&
-\frac{1}{2}
{\cal H}_{\mu\dot\nu\dot\rho}
\Gamma^\mu\Gamma^{\dot\nu\dot\rho}\epsilon'
-\left(\frac{1}{g}+{\cal H}_{\dot1\dot2\dot3}\right)
\Gamma_{\dot1\dot2\dot3}\epsilon'
\nonumber\\&&
-\frac{g^2}{2}\{X^{\dot\mu},X^i,X^j\}
\Gamma^{\dot\mu}\Gamma^{ij}\epsilon'
+\frac{g^2}{6}\{X^i,X^j,X^k\}
\Gamma^{ijk}\Gamma^{\dot1\dot2\dot3}\epsilon',\\
\delta b_{\dot\mu\dot\nu}
&=&-i(\ol\epsilon'\Gamma_{\dot\mu\dot\nu}\Psi'),\\
\delta b_{\mu\dot\nu}
&=&-iV(\ol\epsilon'\Gamma_\mu\Gamma_{\dot\nu}\Psi')
+ig(\ol\epsilon\Gamma_\mu\Gamma_i\Gamma_{\dot1\dot2\dot3}\Psi')
\partial_{\dot\nu}X^i.
\end{eqnarray}

A peculiar property of this SUSY transformation is 
that the perturbative vacuum 
(the configuration with all fields vanishing) 
is not invariant under this transformation
due to the term in $\delta\Psi'$ proportional
to $1/g$.
We can naturally interpret this term as
a contribution of the background $C$-field.
In the M5-brane action coupled to background fields,
the self-dual field strength is defined by
$H=db+C$ (up to coefficients depending on
conventions).
The inclusion of $C$-field in the field strength
is required by
the invariance of the action under $C$-field gauge transformations.
The shift of the field strength ${\cal H}_{\dot1\dot2\dot3}$
by $(1/g)$ in the action as well as in the 
SUSY transformation
suggests that the relation $C\propto g^{-1}$ between 
the Nambu-Poisson structure and the $C$-field background.
This statement of course depends on the normalization of the
gauge field $C$.
For more detail about this relation, see \S\ref{M5D4},
where
we derive the precise form of this relation including the
numerical coefficients.

In fact, M5-brane in a constant $C$-field background
is still $1/2$ BPS.
The effect of the $C$-field is changing
which half of $32$ supersymmetry remain unbroken.
We can find this phenomenon in our six-dimensional theory.
In addition to $16$ supersymmetries we described above,
the theory has $16$ non-linear fermionic symmetries
$\delta^{\rm(nl)}$,
which shift the fermion by a constant spinor
\begin{equation}
\delta^{\rm(nl)}\Psi'=\chi,\quad
\delta^{\rm(nl)}X^i=\delta^{\rm(nl)}b_{\dot\mu\dot\nu}
=\delta^{\rm(nl)}b_{\mu\dot\nu}=0.
\label{chitr}
\end{equation}
The action is invariant under this transformation because
constant functions in $y^{\dot\mu}$ space
are in the center of the $3$-algebra.
The perturbative vacuum is invariant under
the combination of two fermionic symmetries
\begin{equation}
\delta_{\epsilon'}-\frac{1}{g}\delta^{\rm(nl)}_{\epsilon'}.
\end{equation}
In the weak coupling limit $g\rightarrow 0$,
the transformation laws for this combined symmetry
agree with those of
an ${\cal N}=(2,0)$ tensor multiplet
\cite{Romans}.
\begin{eqnarray}
\delta X^i&=&i\ol\epsilon'\Gamma^i\Psi',\\
\delta \Psi'&=&\partial_{\ul\mu}X^i\Gamma^{\ul\mu}\Gamma^i\epsilon'
-\frac{1}{12}H_{\ul\mu\ul\nu\ul\rho}\Gamma^{\ul\mu\ul\nu\ul\rho}\epsilon',\\
\delta b_{\ul\mu\ul\nu}
&=&-i(\ol\epsilon'\Gamma_{\ul\mu\ul\nu}\Psi').
\label{deltabsusy}
\end{eqnarray}
We obtained the transformation (\ref{deltabsusy})
only for $\ul\mu\ul\nu=\dot\mu\dot\nu$ and $\mu\dot\nu$.
To obtain the transformation law of
the $b_{\mu\nu}$ components,
we first compute the transformation
of
${\cal H}_{\dot\mu\dot\nu\dot\rho}$ and ${\cal H}_{\mu\dot\nu\dot\rho}$
by using the transformation law of $b_{\dot\mu\dot\nu}$
and $b_{\mu\dot\nu}$.
Because the field strength is self-dual,
it also gives $\delta \wt{\cal H}_{\mu\nu\dot\rho}$
and $\delta \wt{\cal H}_{\mu\nu\rho}$.
The equations of motion (\ref{maxwell1}) and (\ref{maxwell2})
are the Bianchi identities as well for these components
of field strength.
If we can solve these Bianchi identities
on shell and express them by using $b_{\mu\nu}$,
we can extract the transformation law of $b_{\mu\nu}$
from $\delta \wt{\cal H}_{\mu\nu\dot\rho}$
and $\delta \wt{\cal H}_{\mu\nu\rho}$.
In the free field limit $g=0$,
we can easily carry out this procedure
and obtain (\ref{deltabsusy}) for $b_{\mu\nu}$.

%%%%%%%%%%%%%%%%%%%%%%%%%%%%%%%%%%%%%%%%%%%%%%%%%%%%%%%%%%%%%%%
\section{Derivation of D4 action from M5}
\label{M5D4}

In this section we demonstrate that the double dimensional reduction
of the six-dimensional theory correctly reproduces
the action of non-commutative $U(1)$ gauge theory,
which is realized on a D4-brane in a $B$-field background.

We here recover the overall factor
$T_6$ in the front of the action.
This has mass dimension $6$ and can be regarded as the tension of the
five-brane, while the coupling constant $g$ is a dimensionless parameter.
We should note that this tension $T_6$ is not necessarily the same as the
usual M5-brane tension $T_{\rm M5}$,
because it may be corrected by
the background $C$-field.
We will later determine the parameters $g$ and $T_6$ by
comparing the five dimensional action
obtained by the douple dimensional reduction
of the six-dimensional theory to the non-commutative
$U(1)$ action realized on a D4-brane
in a $B$-field background in type \IIA theory.
Once we obtain the expression for $g$ and $T_6$
in terms of type \IIA parameters, it will be easy to rewrite
them in terms of the M-theory Planck scale and the magnitude
of the $C$-field.

The double dimensional reduction means that we wrap one leg 
of the M5-brane on a compactified dimension, 
so that through Kaluza-Klein reduction 
we get one fewer dimension for both the target space and the world-volume. 
Let us choose the compactified dimension to be $X^{\dot{3}}$. 
In the double dimensional reduction, 
we suppress $y^{\dot{3}}$-dependence of all fields except $X^{\dot{3}}$. 
We have 
\begin{equation}
X^{\dot{3}} =\frac{1}{g}y^{\dot{3}}, \qquad b^{\dot{3}} = 0. 
\label{x3y3}
\end{equation}
We used a gauge symmetry generated by $\Lambda_{\dot1}$ and $\Lambda_{\dot2}$
to set $b^{\dot3}=0$.
We impose the periodicity condition
\begin{equation}
X^{\dot3}\sim X^{\dot3}+L_{11}.
\label{l11}
\end{equation}
The relation (\ref{x3y3}) and (\ref{l11}) implies
that the compactification period of the coordinate $y^{\dot3}$
is $gL_{11}$,
and thus,
the overall factor of
the five dimensional theory becomes
$gL_{11}T_6$.

Let us now first carry out the dimensional reduction 
for the bosonic terms in the action. 
Since all the fields except $X^{\dot{3}}$ have no dependence on $y^{\dot{3}}$, 
we set $\partial_{\dot{3}} = 0$ unless it acts on $X^{\dot{3}}$. 
We will use the notation that indices $\dot{\alpha}, \dot{\beta}, \cdots$ 
take values in $\{\dot{1}, \dot{2}\}$, 
and $a, b, \cdots$ take values in $\{0, 1, 2, \dot{1}, \dot{2}\}$. 
The antisymmetrized tensor $\epsilon^{\dot{\alpha}\dot{\beta}}$ is defined as 
$\epsilon^{\dot{\alpha}\dot{\beta}} = \epsilon^{\dot{\alpha}\dot{\beta}\dot{3}}$.

Expecting that we will obtain a gauge field theory on a D4-brane, 
let us define the gauge potentials 
\begin{equation}
\hat{a}_{\mu}
=b_{\mu\dot{3}}
\qquad 
\hat{a}_{\dot{\alpha}}
= b_{\dot{\alpha}\dot{3}}.
\end{equation}
The covariant derivatives become
\begin{eqnarray}
D_{\mu} X^{\dot{\alpha}}
= -\epsilon^{\dot{\alpha}\dot{\beta}}
\hat{F}_{\mu \dot{\beta}}, \quad 
D_{\mu} X^{\dot{3}} = - \tilde{a}_{\mu}, 
\quad 
D_{\mu} X^i = \hat{D}_{\mu} X^i,
\end{eqnarray}
where $\hat F_{ab}$, $\wt a_\mu$, and $\hat D_a$ are defined by
\begin{eqnarray}
\hat{F}_{ab}
&=& \partial_a \hat{a}_b - \partial_b \hat{a}_a 
+ g\{ \hat{a}_a, \hat{a}_b \},\\
\tilde{a}_{\mu} &=&
\epsilon^{\dot{\alpha}\dot{\beta}} \partial_{\dot{\alpha}}b_{\mu\dot{\beta}},\\
\label{tilde_a}
\hat{D}_\mu \Phi
&=& \partial_\mu \Phi
 + g\{\hat a_\mu, \Phi \}.
\end{eqnarray}
The Poisson bracket $\{\cdot, \cdot\}$ is defined
as the reduction of the Nambu-Poisson bracket
\begin{equation}
\{ f, g \} = \{ y^{\dot{3}}, f, g \}. 
\end{equation}
We note that the components $b_{\mu\dot\beta}$ only show up
through the form $\tilde a_\mu$ in D4 action.
Thus we find that, after double dimensional reduction, 
the scalar kinetic term in the BLG Lagrangian become
\begin{equation}
-\frac{T_6}{2}
\int d^3x\langle
(D_{\mu} X^I)^2 \rangle
=
-\frac{gL_{11}T_6}{2}
\int d^3xd^2y
\left( \tilde{a}_{\mu}^2
+ \hat{F}_{\mu\dot{\alpha}}^2 
+ (\hat{D}_{\mu} X^i)^2 \right).
\end{equation}

The Nambu-Poisson brackets which appear in the potential terms 
of the BLG action are 
\begin{eqnarray}
\{ X^{\dot{1}}, X^{\dot{2}}, X^{\dot{3}} \} 
&=& \frac{1}{g^2} \hat{F}_{\dot{1}\dot{2}} + \frac{1}{g^3}, \\
\{ X^{\dot{3}}, X^{\dot{\alpha}}, X^i \} 
&=& \frac{1}{g^2}\epsilon^{\dot{\alpha}\dot{\beta}} \hat{D}_{\dot{\beta}} X^i, \\
\{ X^{\dot{3}}, X^i, X^j \} 
&=& \frac{1}{g} \{ X^i, X^j \}. 
\end{eqnarray} 
The potential term becomes
\begin{eqnarray}
&&- \frac{T_6}{12}
\int d^3x\langle
g^4 \{ X^I, X^J, X^K \}^2 \rangle
\nonumber\\
&&= 
gL_{11}T_6
\int d^3xd^2y
\left[
-\frac{1}{2}\left(\hat{F}_{\dot{1}\dot{2}} + \frac{1}{g}\right)^2
 - \frac{1}{2} (D_{\dot{\alpha}}X^i)^2 
- \frac{g^2}{4} \{ X^i, X^j \}^2
\right].
\end{eqnarray} 
Upon integration over the base space and removing total derivatives, 
we can replace $(\hat{F}_{\dot{1}\dot{2}} + 1/g)^2$ 
by $\hat{F}_{\dot{1}\dot{2}}^2 + 1/g^2$.

It is also straightforward to show that the Chern-Simons 
term (\ref{CSt}) gets simplified considerably as
\begin{equation}
-\frac{gL_{11}T_6}{2}\int d^3x
d^2y \epsilon^{\mu\nu\lambda} \hat{F}_{\mu\nu} \tilde{a}_{\lambda}. 
\end{equation}
Here again the action depends on $b_{\mu\dot \beta}$
only through $\tilde a_\mu$.
As the action depends on the field $\tilde{a}_{\mu}$ only algebraically
(namely without derivative), we can integrate it out. 
There are only two terms involving $\tilde{a}_{\mu}$ and 
by completing square, 
we find that the effect of integrating out $\tilde{a}_{\mu}$ is 
to replace all terms involving $\tilde{a}_{\mu}$ by 
\begin{equation}
-\frac{gL_{11}T_6}{4}\int d^3xd^2y\hat{F}_{\mu\nu}^2.
\end{equation}

The fermion part
can be evaluated similarly.
The covariant derivatives and bracket are
\ba
&& \Gamma^\mu D_\mu \Psi' = \Gamma^\mu(\partial_\mu \Psi'
+g\left\{ a_\mu, \Psi'\right\}):=\Gamma^\mu\hat{D}_\mu \Psi' \,,\\
&& \frac{1}{2}\Gamma_{IJ}\{X^I, X^J, \Psi'\}=\Gamma_{\dot\alpha}\Gamma_{\dot1\dot2\dot3}
\hat D_{\dot\beta}\Psi'+
\Gamma_{\dot3}\Gamma_i\left\{X^i,\Psi'\right\}\,,\\
&&\hat D_{\dot\beta}\Psi':=
\partial_{\dot\beta}\Psi'+g\{a_{\dot\beta},\Psi'\}.
\ea

It is quite remarkable that, after collecting all the kinetic, potential
and Chern-Simons temrs,
the 4+1 dimensional Lorentz invariance 
is restored
(up to the breaking by the non-commutativity). 
The sum of all these terms is simply 
\begin{eqnarray}
&&gL_{11}T_6\int d^3xd^2y
\left[
- \frac{1}{2}(\hat{D}_a X^i)^2
 - \frac{1}{4}\hat F_{ab}^2
- \frac{g^2}{4} \{ X^i, X^j \}^2
- \frac{1}{2g^2}
\right.
\nn\\
&&\hspace{3cm}
\left.
+\frac{i}{2}\left(
\bar\Psi'' \Gamma^a\hat D_a\Psi''
+g\bar\Psi''\Gamma_i
\{X^i,\Psi''\}
\right)
\right].
\label{LD4}
\end{eqnarray}
We performed the unitary transformation
$\Psi'=(1/\sqrt2)(\Gamma_{\dot3}+\Gamma^7)\Psi''$
to obtain the correct chirality condition $\Gamma_{\dot 3}\Psi''=-\Psi''$
for the gaugino on the D4-brane.
(Nore that $\dot 3$ is now the ``eleventh'' direction
and $\Gamma_{\dot3}$ is the chirality matrix in \IIA theory.)

Let us compare the action (\ref{LD4}) with the known result\cite{Aoki:1999vr,SeibergWitten}
for a D4-brane in a
$B$-field background,
and match the parameters in this theory and
those in type \IIA string theory.
The non-commutative gauge theory on D4-brane in
a $B$-field background is described with the Moyal product $*$,
and the corresponding commutator, 
the so-called Moyal bracket $[\cdot,\cdot]_{\rm Moyal}$,
defined by
\begin{equation}
f(x)*g(x)=
e^{\frac{i}{2}\theta^{ij}
\frac{\partial}{\partial\xi^i}
\frac{\partial}{\partial\zeta^j}}
f(x+\xi)g(x+\zeta)|_{\xi=\zeta=0},
\end{equation}
\begin{equation}
[f,g]_{\rm Moyal}
=f*g-g*f
=\theta^{ij}\partial_if\partial_jg+{\cal O}(\theta^3).
\end{equation}
The non-commutativity parameter $\theta^{ij}$
has the dimension of (length)$^2$.
Because the action (\ref{LD4}) includes
only finite powers of derivatives, it should
be compared to the weak coupling limit $\theta\rightarrow 0$
of the non-commutative gauge theory.
These two match if we truncate the Moyal bracket
into the Poisson bracket by
\begin{equation}
[f,g]_{\rm Moyal}\rightarrow \frac{\theta}{T_{\rm str}}\{f,g\},
\end{equation}
where we turn on the non-commutativity in the $\dot1$-$\dot2$ directions
by setting
\begin{equation}
\theta^{\dot1\dot2}=\frac{\theta}{T_{\rm str}},\quad
\theta^{\mu\dot\alpha}=\theta^{\mu\nu}=0.
\end{equation}
Note that $\theta$ is defined as a dimensionless parameter.
In the small $\theta$ limit,
the bosonic part of the action of the non-commutative $U(1)$ gauge theory
on a D4-brane is given by\cite{Aoki:1999vr,SeibergWitten}
\begin{eqnarray}
S
=\frac{T_{D4}}{\theta}\int d^3xd^2y\left[
-\frac{1}{2}(D_a X^i)^2
-\frac{1}{4T_{\rm str}}F_{ab}^2
-\frac{\theta^2}{4}\{X^i,X^j\}^2
-\frac{1}{2\theta^2}
\right],
\label{iso}
\end{eqnarray}
in the open string frame.
The world-volume coordinate $y^{\dot\alpha}$
in the open string frame is related to the target space coordinates
$X^{\dot\alpha}$ by
\begin{equation}
X^{\dot\alpha}=\frac{1}{\theta} y^{\dot\alpha}.
\label{iiaxy}
\end{equation}
The covariant
derivative and the field strength are
\begin{equation}
D_a X^i=\partial_\mu X^i+\frac{\theta}{T_{\rm str}}\{A_a,X^i\},\quad
F_{ab}=\partial_a A_b-\partial_b A_a+\frac{\theta}{ T_{\rm str}}\{A_a,A_b\}.
\end{equation}
We normalize the gauge field $A_a$ so that it couples to 
the string endpoints by charge $1$ through
the boundary coupling $S=\int_{\partial{\rm F1}} A$ of
the fundamental string world-sheet,
and this gauge field has mass dimension $1$.
In the weak coupling limit, the non-commutativity parameter $\theta$ is
related to the background $B$-field by
\begin{equation}
B
=T_{\rm str}\theta dX^{\dot1}\wedge dX^{\dot2}
=\frac{T_{\rm str}}{\theta}dy^{\dot1}\wedge dy^{\dot2}.
\label{bfield}
\end{equation}

By comparing two actions (\ref{LD4}) and (\ref{iso}),
we obtain the following relations among parameters:
\begin{eqnarray}
T_6&=&\frac{T_{M5}}{\theta^2},\label{t6}\\
g&=&\theta.\label{gc}
\end{eqnarray}
To relate quantities in \IIA and M-theory,
we use the following relations among
tensions of M-branes and \IIA-branes.
\begin{equation}
T_{D4} = L_{11} T_{M5}, \quad
T_{\rm str}=L_{11}T_{D2} =L_{11}T_{M2}.
\end{equation}
The relation $T_{M2}^2=2\pi T_{M5}$ is also useful.

In addition to the agreement of the action through the relations
(\ref{t6}) and (\ref{gc}),
we can check the consistency in some places.

Firstly,
the relation (\ref{iiaxy}) between the world-volume coordinates
and the target space coordinates can naturally be lifted to
the relation (\ref{xeqyb}).

Secondly, the overall factor $T_6$ agrees with the effective
tension of M2-branes induced by the background $C$-field.
The background $B$-field (\ref{bfield}) is lifted to the
background three-form field
\begin{equation}
C_3
=\theta T_{\rm M2} dX^{\dot1}\wedge dX^{\dot2}\wedge dX^{\dot3}
=\frac{T_{\rm M2}}{\theta^2}dy^{\dot1}\wedge dy^{\dot2}\wedge dy^{\dot3}.
\label{backc3}
\end{equation}
(We use the convention in which the gauge fields $B$ and $C$ couple
to the world-volume of corresponding branes
by charge $1$ through the couplings $\int_{\rm F1}B$ and $\int_{\rm M2}C$.)
Each flux quantum of this background field induces the charge of a single
M2-brane on the M5-brane,
and effective M2-brane density in the $y$-space is
$\theta^{-2}T_{\rm M2}/(2\pi)$.
Thus, if we assume that the tension of M5-brane is dominated by the
induced M2-branes, the effective tension becomes
$T_{\rm M2}\times \theta^{-2}T_{\rm M2}/(2\pi)=\theta^{-2}T_{\rm M5}$.
This agrees with the overall coefficients
$T_6$ given in the relation (\ref{t6}).

Finally, the charge of the self-dual strings is
consistent with the Dirac's quantization condition.
From the comparison of the actions
we obtain the relation of gauge fields
\begin{equation}
\hat a_a=\frac{1}{T_{\rm str}}A_a.
\label{aA}
\end{equation}
As we mentioned above, the gauge field $A$ couples to string endpoints
by charge $1$.
By the correspondence (\ref{aA}) we can determine the
strength of the coupling of $\hat a$ and $b$ to
boundaries of the corresponding branes.
The boundary interactions are given by
\begin{equation}
S
=T_{\rm str}\int_{\partial{\rm F1}}\hat a
=\frac{T_{\rm M2}}{\theta}\int_{\partial{\rm M2}} b .
\label{sab}
\end{equation}
To obtain the second equality in
(\ref{sab}), we used
the fact that a string endpoint is lifted
to an M2-brane boundary wrapped on the $S^1$
along $y^{\dot3}$ with period $gL_{11}$.
The coupling (\ref{sab}) shows that
the charge of self-dual strings
(boundary of M2-branes ending on the M5-brane)
is $Q=\theta^{-1} T_{\rm M2}$.
Because the gauge field $b$ is a self-dual
field, $Q$ is the electric charge as well as
the magnetic charge of a self-dual string,
and it must satisfy the Dirac's quantization condition
\begin{equation}
\frac{Q^2}{T_6}=2\pi.
\end{equation}
We can easily check that this relation certainly holds.

We can now explain the constant shift in the
field strength as follows.
The M2-brane action includes the following coupling to the
bulk $3$-form field $C$ and the self-dual $2$-form field $b$:
\begin{equation}
S_{\rm M2}=\int_{\rm M2} C_3+\frac{T_{\rm M2}}{\theta}\int_{\partial{\rm M2}}b.
\end{equation}
The gauge invariance of this action requires that under the gauge transformation
$\delta C_3=d\alpha_2$, the self-dual field on the M5-brane must transform
as $\delta b_2=-\alpha_2/(\theta^{-1} T_{\rm M2})$.
Thus, the gauge invariant field strength $H$ of the tensor field
$b$ should be defined by
\begin{equation}
H=db+\frac{\theta}{T_{M2}}C.
\end{equation}
Therefore, the background gauge field 
(\ref{backc3}) shifts the field strength as
\begin{equation}
H=db+\frac{1}{\theta} dy^{\dot1}\wedge dy^{\dot2}\wedge dy^{\dot3}.
\end{equation}
This is the same as the constant shift in the definition (\ref{h30def})
of ${\cal H}_{\dot1\dot2\dot3}$.

Now we have relations between parameters in
the BLG theory and those in M-theory.
The D4-brane action obtained by the double dimensional reduction
is the weak coupling ($g=\theta\rightarrow0$)
limit of non-commutative $U(1)$ theory
because the Moyal bracket is replaced by the Poisson bracket.
The coupling constant is determined by the background
$C$-field, and the weak coupling means strong $C$-field background
through the relation (\ref{backc3}).
Our M5-brane theory is expected to apply better 
to the limit of large $C$-field background. 
This is also confirmed in the comparison of the five-brane tension.
As we mentioned above, the effective tension $T_6$ is dominated
by the tension of M2-branes induced by the background $C$-field.
This is the case when the background $C$-field is very large.

For a finite $C$-field background, we expect that 
the Nambu-Poisson bracket should be replaced 
by a quantum Nambu bracket.

%%%%%%%%%%%%%%%%%%%%%%%%%%%%%%%%%%%%%%%%%%%%%%%%%%%%%%%%
\section{Seiberg-Witten map}

It was found by Seiberg and Witten \cite{SeibergWitten}
that the gauge symmetry on a noncommmutative space 
can be matched with the gauge symmetry on a classical space 
via the so-called Seiberg-Witten map 
\begin{equation} \label{SW}
\hat{\delta}_{\hat{\lambda}} \hat{\Phi}(\Phi) 
= \hat{\Phi}(\Phi + \delta_{\lambda}\Phi) - \hat{\Phi}(\Phi), 
\end{equation} 
where $\hat{\Phi}(\Phi)$ is the field variable, 
and $\hat{\delta}_{\hat{\lambda}}$ the gauge transformation 
in the noncommutative gauge theory corresponding 
to $\Phi$ and $\delta_{\lambda}$ 
living on the classical space. 
The gauge transformation parameter 
$\hat{\lambda}(A, \lambda)$ 
in the noncommutative gauge theory is 
a function of the gauge potential $A$ and 
gauge transformation parameter $\lambda$ 
in the gauge theory on classical spacetime. 
The Seiberg-Witten map is found as an infinite expansion 
of the noncommutativity parameters. 

In this section we find the Seiberg-Witten map 
connecting the gauge theories on spacetimes 
with and without the Nambu-Poisson structure, 
corresponding to M5-brane theories in trivial 
or constant $C$-field background.  
In this section only, we denote all variables 
in our M5-brane theory by symbols with hats, 
and those in trivial backgrounds by symbols without hats. 
As $g\rightarrow 0$, 
the variables with hats should reduce to 
those without hats.

In the trivial background, we have 
the gauge fields $b_{\dot{\mu}\dot{\nu}}$, $b_{\mu\dot{\mu}}$ 
and gauge transformations
\begin{equation}
\delta_{\Lambda} b_{\dot{\mu}\dot{\nu}} = 
\partial_{\dot{\mu}} \Lambda_{\dot{\nu}} 
- \partial_{\dot{\nu}} \Lambda_{\dot{\mu}}, \qquad 
\delta_{\Lambda} b_{\mu\dot{\mu}} = 
\partial_{\mu} \Lambda_{\dot{\mu}} 
- \partial_{\dot{\mu}} \Lambda_{\mu}. \label{gtb0}
\end{equation}
In the M5-brane theory with a $C$-field background, 
we have 
\begin{eqnarray}
\hat{\delta}_{\hat \Lambda}\hat{b}_{\dot{\mu}\dot{\nu}}
&=&\partial_{\dot\mu}\hat{\Lambda}_{\dot\nu}
-\partial_{\dot\nu}\hat{\Lambda}_{\dot\mu}
+g\hat{\kappa}^{\dot\lambda}
\partial_{\dot\lambda}  \hat{b}_{\dot{\mu}\dot{\nu}}, \label{gtb1} \\
\hat{\delta}_{\hat\Lambda}\hat{b}_{\mu\dot{\mu}}
&=&\partial_\mu\hat{\Lambda}_{\dot\mu}
-\partial_{\dot\mu}\hat{\Lambda}_{\mu}
+g\hat{\kappa}^{\dot\nu}
\partial_{\dot\nu}\hat{b}_{\mu\dot\mu}
+g(\partial_{\dot\mu}\hat{\kappa}^{\dot\nu})
\hat{b}_{\mu\dot\nu}. \label{gtb2}
\end{eqnarray}

It will be convenient to use the following variables 
\begin{equation}
\hat{b}^{\dot\mu} \equiv 
\epsilon^{\dot\mu\dot\nu\dot\lambda} 
\hat{b}_{\dot\nu\dot\lambda}, \qquad 
\hat{B}_{\mu}{}^{\dot\mu} \equiv
\epsilon^{\dot\mu\dot\nu\dot\lambda} 
\partial_{\dot\nu}\hat{b}_{\mu\dot\lambda}. 
\end{equation}
instead of $\hat{b}_{\dot\mu\dot\nu}$ and $\hat{b}_{\mu\dot{\mu}}$. 
Similarly we define $b^{\dot\mu}$ and $B_{\mu}{}^{\dot{\mu}}$ 
in the same way. 
We shall impose the constraints  
\begin{equation} \label{dB=0}
\partial_{\dot\mu} \hat{B}_{\mu}{}^{\dot\mu} = 0, 
\qquad \mbox{and} \qquad
\partial_{\dot\mu} B_{\mu}{}^{\dot\mu} = 0
\end{equation}
on $\hat{B}_{\mu}{}^{\dot\mu}$ and $B_{\mu}{}^{\dot\mu}$
so that the existence of $\hat{b}_{\mu\dot\mu}$ 
and $b_{\mu\dot\mu}$ is guaranteed when the former are given. 
The gauge transformations of the new variables are 
\begin{equation}
\hat{\delta}_{\hat{\Lambda}} \hat{b}^{\dot\mu} 
= \hat{\kappa}^{\dot\mu} 
+ g\hat{\kappa}^{\dot\nu}\partial_{\dot\nu}\hat{b}^{\dot\mu}, 
\qquad 
\hat{\delta}_{\hat{\Lambda}} \hat{B}_{\mu}{}^{\dot\mu} 
= \partial_{\mu}\hat{\kappa}^{\dot\mu}
+ g\hat{\kappa}^{\dot\nu}\partial_{\dot\nu}\hat{B}_{\mu}{}^{\dot\mu}
- g(\partial_{\dot\nu}\hat{\kappa}^{\dot\mu})\hat{B}_{\mu}{}^{\dot\nu}, 
\end{equation}
where 
\begin{equation}
\hat{\kappa}^{\dot\mu} \equiv \epsilon^{\dot\mu\dot\nu\dot\lambda}
\partial_{\dot\nu} \hat{\Lambda}_{\dot\lambda}
\end{equation}
was denoted by $\delta_\Lambda y^{\dot\mu}$ above, and it satisfies 
\begin{equation} \label{dk=0}
\partial_{\dot\mu} \hat{\kappa}^{\dot\mu} = 0.
\end{equation}
Analogous to all these equations above, 
we have the corresponding equations for variables 
on a M5-brane in the trivial background. 
They can be obtained by taking the $g\rightarrow 0$ limit as 
\begin{eqnarray}
\delta_{\Lambda} b^{\dot\mu} 
= \kappa^{\dot\mu}, 
\qquad 
\delta_{\Lambda} B_{\mu}{}^{\dot\mu} 
= \partial_{\mu}\kappa^{\dot\mu}, 
\qquad 
\kappa^{\dot\mu} = \epsilon^{\dot\mu\dot\nu\dot\lambda}
\partial_{\dot\nu} \Lambda_{\dot\lambda}. 
\end{eqnarray}

We need two Seiberg-Witten maps $\hat{b}^{\dot\mu}(b)$ 
and $\hat{B}_{\mu}{}^{\dot\mu}(B, b)$ satisfying 
\begin{eqnarray}
\hat{\delta}_{\hat{\Lambda}} \hat{b}^{\dot\mu}(b)
&=& \hat{b}^{\dot\mu}(b + \delta_{\Lambda}b) - \hat{b}^{\dot\mu}(b), \label{SWcondb}
\\
\hat{\delta}_{\hat{\Lambda}} \hat{B}_{\mu}{}^{\dot\mu}(B, b)
&=& \hat{B}_{\mu}{}^{\dot\mu}
(B + \delta_{\Lambda} B, b + \delta_{\Lambda} b) 
- \hat{B}_{\mu}{}^{\dot\mu}(B, b).  
\label{SWcondB}
\end{eqnarray}
The solutions would be infinite expansions in $g$. 
To the first order terms in $g$, we find the solution as 
\begin{eqnarray}
\hat{b}^{\dot\mu}(b) &=& b^{\dot\mu} + 
\frac{g}{2} b^{\dot\nu}\partial_{\dot\nu} b^{\dot\mu} + 
\frac{g}{2} b^{\dot\mu} \partial_{\dot\nu} b^{\dot\nu} 
+ {\cal O}(g^2), 
\label{SWmb} \\
\hat{B}_{\mu}{}^{\dot\mu}(B, b) &=& B_{\mu}{}^{\dot\mu} + 
g b^{\dot\nu} \partial_{\dot\nu} B_{\mu}{}^{\dot\mu}  
- \frac{g}{2} b^{\dot\nu} \partial_{\mu}\partial_{\dot\nu} 
b^{\dot\mu} + \frac{g}{2} b^{\dot\mu} 
\partial_{\mu}\partial_{\dot\nu} b^{\dot\nu} 
+ g \partial_{\dot\nu} b^{\dot\nu} B_{\mu}{}^{\dot\mu} 
\nonumber \\
&&
- g \partial_{\dot\nu} b^{\dot\mu} B_{\mu}{}^{\dot\nu} 
- \frac{g}{2} \partial_{\dot\nu} b^{\dot\nu} 
\partial_{\mu} b^{\dot\mu} 
+ \frac{g}{2} \partial_{\dot\nu} b^{\dot\mu} 
\partial_{\mu} b^{\dot\nu} 
+ {\cal O}(g^2), 
\label{SWmB} \\ 
\hat{\kappa}^{\dot\mu}(\kappa, b) &=& \kappa^{\dot\mu} + 
\frac{g}{2} b^{\dot\nu}\partial_{\dot\nu} \kappa^{\dot\mu} 
+ \frac{g}{2} (\partial_{\dot\nu} b^{\dot\nu}) \kappa^{\dot\mu} 
- \frac{g}{2} (\partial_{\dot\nu} b^{\dot\mu}) \kappa^{\dot\nu} 
+ {\cal O}(g^2). \label{SWmk}
\end{eqnarray}
Some of the coefficients here are not 
completely fixed by the Seiberg-Witten map conditions 
(\ref{SWcondb}-\ref{SWcondB}), 
but they are uniquely determined by 
the requirement that 
the constraint (\ref{dB=0}) is preserved 
by the Seiberg-Witten map. 
It should be possible to solve for higher order terms order by order. 

To be complete, let us consider $X^I$ and $\Psi$, 
or anything that transforms like 
\begin{equation}
\delta \hat{\Phi} = g\hat{\kappa}^{\dot{\mu}} 
\partial_{\dot\mu} \hat{\Phi}. 
\end{equation}
The classical counterpart of $\hat{\Phi}$ has 
\begin{equation}
\delta\Phi = 0. 
\end{equation}
The Seiberg-Witten map condition (\ref{SW}) 
is easy to solve for this case 
to obtain the first order terms, 
and the solution is 
\begin{equation}
\hat{\Phi} = \Phi + g b^{\dot{\mu}} \partial_{\dot\mu}\Phi 
+ {\cal O}(g^2). 
\end{equation}

A comment is needed here regarding the map between 
$\hat{\kappa}^{\dot\mu}$ and $\kappa^{\dot\mu}$, 
which are defined in terms of the gauge transformation 
parameters $\hat{\Lambda}_{\dot\mu}$ and $\Lambda_{\dot\mu}$. 
If one wants to determine the map between 
$\hat{\Lambda}_{\dot\mu}$ and $\Lambda_{\dot\mu}$, 
it is necessary to fix the ambiguity in the gauge parameters 
for a given gauge transformation. 
It is obvious that the transformation of the gauge parameters 
\begin{eqnarray}
&\hat{\Lambda}_{\dot\mu} \rightarrow \hat{\Lambda}_{\dot\mu} 
+ \partial_{\dot\mu} \hat{\xi}, \qquad
\hat{\Lambda}_{\mu} \rightarrow \hat{\Lambda}_{\mu} 
+ \partial_{\mu} \hat{\xi}, \\ 
&\Lambda_{\dot\mu} \rightarrow \Lambda_{\dot\mu} 
+ \partial_{\dot\mu} \xi, \qquad
\Lambda_{\mu} \rightarrow \Lambda_{\mu} 
+ \partial_{\mu} \xi 
\end{eqnarray}
does not change the gauge transformations
(\ref{gtb0}-\ref{gtb2}) at all. 
To avoid this ambiguity in the gauge transformation parameters, 
we can use $\hat{\kappa}^{\dot\mu}$ and $\kappa^{\dot\mu}$ instead, 
and the existence of $\hat{\Lambda}_{\dot\mu}$ and $\Lambda_{\dot\mu}$ 
are guaranteed by the constraints (\ref{dk=0}). 
One can check that the constraint (\ref{dk=0}) is 
preserved by the Seiberg-Witten map (\ref{SWmk}). 

The ambiguity involved here is in the same form 
as a gauge transformation of 1-form gauge fields, 
and hence we can ``gauge fix'' the gauge transformation 
parameters by the following constraints 
\begin{equation}
\partial_{\dot\mu} \hat{\Lambda}^{\dot\mu} = 0, \qquad 
\partial_{\dot\mu} \Lambda^{\dot\mu} = 0. 
\end{equation}
(Here we are raising and lowering indices using 
the metric $\delta_{\dot\mu\dot\nu}$ on ${\cal N}$.) 
We can solve these constraints by 
\begin{equation}
\hat{\Lambda}^{\dot\mu} = \epsilon^{\dot\mu\dot\nu\dot\lambda}
\partial_{\dot\nu} \hat{\eta}_{\dot\lambda}, \qquad
\Lambda^{\dot\mu} = \epsilon^{\dot\mu\dot\nu\dot\lambda}
\partial_{\dot\nu} \eta_{\dot\lambda},
\end{equation}
and again we can demand the constraints 
\begin{equation}
\partial_{\dot\mu} \hat{\eta}^{\dot\mu} = 0, \qquad 
\partial_{\dot\mu} \eta^{\dot\mu} = 0 
\end{equation} 
on the parameters $\hat{\eta}^{\dot\mu}$ and $\eta^{\dot\mu}$, 
in terms of which we have 
\begin{equation}
\hat{\kappa}^{\dot\mu} = - \partial^2 \hat{\eta}^{\dot\mu}, \qquad 
\kappa^{\dot\mu} = - \partial^2 \eta^{\dot\mu}, 
\end{equation} 
where $\partial^2 \equiv \partial_{\dot\mu}\partial^{\dot\mu}$
is the Laplace operator on ${\cal N}$. 
This allows us to deduce the Seiberg-Witten map between 
$\hat{\eta}^{\dot\mu}$ and $\eta^{\dot\mu}$, 
and finally the Seiberg-Witten map for the gauge transformation 
parameters can be expressed in the following nonlocal form
\begin{equation}
\hat{\Lambda}^{\dot\mu} = \Lambda^{\dot\mu} 
- \frac{g}{2} \epsilon^{\dot\mu\dot\nu\dot\lambda} \partial^{-2}
\partial_{\dot\nu} [
b^{\dot\rho}\partial_{\dot\rho}\kappa_{\lambda}
+(\partial_{\dot\rho}b^{\dot\rho})\kappa_{\lambda} 
-(\partial_{\dot\rho}b_{\dot\lambda})\kappa^{\dot{\rho}}
] + \cdots.
\end{equation}

%%%%%%%%%%%
%To be put in sec. 4.3
%
%Note that, although the SUSY transformation laws 
%of the deformed M5 theory coincides with those 
%for the M5 in trivial background with the identification 
%$\hat{H}_{\dot{1}\dot{2}\dot{3}} 
%= H_{\dot{1}\dot{2}\dot{3}} + g^{-1}$ 
%in the limit $g\rightarrow 0$, 
%we can not immediately conclude that $g = C^{-1}$, 
%because the infinite $C$-field limit is not 
%the same as the trivial background. 

%%%%%%%%%%%%%%%%%%%%%%%%%%%%%%%%%%%%%%%%%%%%%%%%%%%%%%%%%%%%%%%%

\section{Interpretations of the M5-brane theory}
\label{geometry}

\subsection{As a field theory of 
the Nambu-Poisson structure}

The gauge symmetry of the M5 world-volume theory 
is the volume-preserving diffeomorphism on ${\cal N}$. 
The transformation law for both $X^i$ and $\Psi$ are 
given in the same form
\begin{equation} \label{gt}
\delta_{\Lambda} \Phi = 
g \delta_\Lambda y^{\dot{\mu}} \partial_{\dot{\mu}} \Phi, 
\end{equation}
where the volume-preserving coordinate transformation
\begin{equation}
\delta_\Lambda y^{\dot{\mu}} = \epsilon^{\dot{\mu}\dot{\nu}\dot{\lambda}}
\partial_{\dot{\nu}}\Lambda_{\dot{\lambda}} 
\end{equation}
is parametrized by three arbitrary functions $\Lambda_{\dot\mu}$. 

In the above we have considered $b_{\mu\dot{\mu}}$ 
and $b_{\dot{\mu}\dot{\nu}}$ as the gauge fields for 
the volume-preserving diffeomorphisms.
Here we give a geometrical interpretation to these quantities 
in terms of deformations of the Nambu-Poisson structure. 

The degrees of freedom corresponding to 
$b_{\dot{\mu}\dot{\nu}}$, or equivalently $b^{\dot{\mu}}$, 
is easy to understand.
It arises in $X^{\dot{\mu}}$ (\ref{xeqyb}), 
and $y^{\dot\mu}$ is a longitudinal coordinate 
on the $M_5$ brane. 
Thus $b^{\dot{\mu}}$ corresponds to 
a coordinate transformation on ${\cal N}$, 
which is not necessarily volume-preserving 
because $\partial_{\dot{\mu}} b^{\dot{\mu}}$ 
may be nonzero. 
In fact, one can view $b^{\dot{\mu}}$ as 
a parametrization of the deformations of 
the Nambu-Poisson structure due to 
a change of the coordinates 
\begin{equation}
y^{\dot{\mu}} \quad \rightarrow \quad 
gX^{\dot{\mu}} = y^{\dot{\mu}} + g b^{\dot{\mu}}. 
\end{equation}
That is, 
\begin{equation}
\{ f, g, h \} = 
\epsilon^{\dot{\mu}\dot{\nu}\dot{\lambda}} 
\partial_{\dot{\mu}} f \partial_{\dot{\nu}} g 
\partial_{\dot{\lambda}} h 
\quad \rightarrow \quad 
g^{-3} \epsilon^{\dot{\mu}\dot{\nu}\dot{\lambda}} 
\frac{\partial}{\partial X^{\dot{\mu}}} f 
\frac{\partial}{\partial X^{\dot{\nu}}} g 
\frac{\partial}{\partial X^{\dot{\lambda}}} h. 
\end{equation}
While $b^{\dot{\mu}}$ is used to parametrize 
deformations of the Nambu-Poisson structure, 
infinitesimal coordinate transformations 
\begin{equation}
\delta b^{\dot{\mu}} = \delta y^{\dot{\mu}} 
+ g\delta y^{\dot{\nu}}\partial_{\dot{\nu}} b^{\dot{\mu}} 
\stackrel{g\rightarrow 0}{\rightarrow} \delta y^{\dot{\mu}} ,
\end{equation} 
which preserve 
the Nambu-Poisson bracket, should be regarded 
as gauge transformations

The other gauge potential $b_{\mu\dot{\mu}}$ 
appears in the covariant derivative (\ref{dmu})
\begin{equation} \label{DB}
D_{\mu} = \partial_{\mu} - 
g B_{\mu}{}^{\dot{\mu}} \partial_{\dot{\mu}}, 
\end{equation}
where 
\begin{equation} \label{Bb}
B_{\mu}{}^{\dot{\mu}} =
 \epsilon^{\dot{\mu}\dot{\nu}\dot{\lambda}}
\partial_{\dot{\nu}} b_{\mu\dot{\lambda}}.
\end{equation} 
Formally, the expression of the $D_{\mu}$ 
suggests that $B_{\mu}{}^{\dot{\mu}}$ is 
the gauge potential and $\partial_{\dot{\mu}}$ 
is the gauge symmetry generator, 
which also appears in (\ref{gt}). 
Indeed, instead of defining $B_{\mu}{}^{\dot{\mu}}$ 
in terms of $b_{\mu\dot{\mu}}$, 
one can view $B_{\mu}{}^{\dot{\mu}}$ 
as the fundamental gauge potential, 
and guarantee the existence of $b_{\mu\dot{\mu}}$ 
through the constraint 
\begin{equation} \label{constraintB}
\partial_{\dot{\mu}} B_{\mu}{}^{\dot{\mu}} = 0.
\end{equation}

The gauge field $B_{\mu}{}^{\dot{\mu}}$ is 
reminiscent of the gauge field parametrizing 
complex structure deformations on a Calabi-Yau 
3-manifold in the Kodaira-Spencer theory \cite{BCOV}. 

Consider a generic 6 dimensional space 
equipped with a Nambu-Poisson structure. 
The decomposability of the Nambu-Poisson bracket 
implies that locally we can always choose 3 coordinates 
$y^{\dot{\mu}}$ such that the Nambu-Poisson bracket 
is just the Jacobian factor (\ref{NPB}). 
Thus the Nambu-Poisson structure induces 
the separation of local coordinates into 
the two sets $\{x^{\mu}\}$ and $\{ y^{\dot{\mu}}\}$. 
This is analogous to the situation of a complex manifold, 
for which there are holomorphic $z^i$ 
and anti-holomorphic $\bar{z}^{\bar{i}}$ coordinates. 
A deformation of the complex structure can 
be described by specifying how the notion of holomorphicity 
is changed. 
A function on the complex manifold is holomorphic if 
\begin{equation}
\bar{\partial} f = 0 \rightarrow \partial_{\bar{i}} f = 0. 
\end{equation}
When the complex structure is deformed, 
the anti-holomorphic exterior derivative is changed 
\begin{equation} \label{ahed}
\bar{\partial} \rightarrow \bar{\partial}_A 
= d\bar{z}^{\bar{i}}D_{\bar{i}}, 
\end{equation}
where 
\begin{equation}
D_{\bar{i}} = \partial_{\bar{i}} + A_{\bar{i}}{}^i \partial_i. 
\end{equation}
The gauge potential $A_{\bar{i}}{}^i$ 
parametrizes how much mixing occurs between 
the holomorphic and anti-holomorphic coordinates 
due to the deformation of complex structure. 
In our M5-brane theory, 
$B_{\mu}{}^{\dot{\mu}}$ plays a similar role 
as $A_{\bar{i}}{}^i$, 
and the covariant derivative $D_{\mu}$ 
can be viewed as a deformation 
of the derivative with respect to $x^{\mu}$. 
It should be understood as a specification 
of how much the coordinates $x^{\mu}$ and $y^{\dot{\mu}}$ 
are mixed by a deformation of the Nambu-Poisson structure. 
While the Kodaira-Spencer theory \cite{BCOV}
is a dynamical theory of the complex structure, 
the M5-brane theory 
can be understood as a dynamical theory of 
the Nambu-Poisson structure. 

To be more persuasive, we can make the analogy 
between the Kodaira-Spencer theory of complex structure 
and the M5-brane theory of Nambu-Poisson structure 
more explicit. 
For a Calabi-Yau 3-fold, 
there is a unique holomorphic $(3, 0)$-form 
$\Omega = \frac{1}{3!} \Omega_{ijk} dz^i dz^j dz^k$.
This allows us to impose a constraint 
on the gauge potential $A$ as 
\begin{equation}
\label{constraintA}
\partial A' = 0 \Leftrightarrow 
\Omega^{ijk} \partial_i A_{\bar{i}jk} = 0, 
\end{equation} 
where 
\begin{equation}
A' = (\Omega\cdot A), \qquad \mbox{i.e.} \qquad
A'_{\bar{i}jk} = \Omega_{ijk} A_{\bar{i}}{}^i. 
\end{equation}
The $(3, 0)$-form $\Omega$ is analogous to 
the Nambu-Poisson tensor field, 
which is taken to be 
$g \epsilon^{\dot{\mu}\dot{\nu}\dot{\lambda}}$ 
by suitably choosing the coordinates $y^{\dot{\mu}}$. 
If we carry out the substitution
\begin{equation} \label{replace}
A\rightarrow B, \qquad 
\Omega_{ijk} \rightarrow 
g \epsilon_{\dot{\mu}\dot{\nu}\dot{\lambda}}, \qquad 
z^i \rightarrow y^{\dot{\mu}},
\end{equation} 
the constraint (\ref{constraintA}) is precisely 
the constraint (\ref{constraintB}) which guarantees 
the existence of $b_{\mu\dot{\mu}}$. 

Furthermore, 
the Kodaira-Spencer equation \cite{KS}, 
which is equivalent to the nilpotency condition 
of the deformed anti-holomorphic exterior derivative 
(\ref{ahed}), is 
\begin{equation} \label{KSeq}
\bar{\partial}_A \bar{\partial}_A = 0, \quad
\Leftrightarrow \quad 
[ D_{\bar{i}}, D_{\bar{j}} ] = 0. 	
%\bar{\partial}A + \frac{1}{2} [A, A] = 0, 
\end{equation}
%where
%\begin{equation}
%\bar{\partial}_A = 
%\bar{\partial} + A, \qquad
%A = d\bar{z}^{\bar{i}} A_{\bar{i}}{}^i \partial_i . 
%\end{equation}

If we turn off all other fields $X^i, \Psi$ and $b^{\dot{\mu}}$, 
the equation of motion for $B_{\mu}{}^{\dot{\mu}}$ is 
\begin{equation} \label{DD}
[D_{\mu}, D_{\nu}] = 0.
\end{equation}
This is exactly what one obtains from (\ref{KSeq}) 
via the replacement (\ref{replace}). 

To summarize, 
while $b_{\mu\dot{\mu}}$ and $b_{\dot{\mu}\dot{\nu}}$ 
are viewed as the gauge potentials 
for the gauge symmetry of coordinate transformations 
preserving a given Nambu-Poisson structure, 
$B_{\mu}{}^{\dot{\mu}}$ and $b^{\dot{\mu}}$ 
should be viewed as two types of deformation parameters 
of the Nambu-Poisson structure of the M5-brane world-volume. 
We have $b^{\dot{\mu}}$ specifying the change of 
the Nambu-Poisson structure due to 
a change of coordinates $\delta y^{\dot{\mu}}$ in ${\cal N}$ 
(so that the volume form is changed), 
and $B_{\mu}{}^{\dot{\mu}}$ the change due to 
a mixing of the two classes of coordinates 
$x^{\mu}$ and $y^{\dot{\mu}}$. 
The gauge symmetry corresponds to redundant 
descriptions of deformations of the Nambu-Poisson structure. 
The M5-brane theory with a self-dual gauge field
can thus be interpreted as 
a dynamical theory of the Nambu-Poisson structure.

\subsection{As an effective theory in large $C$-field background}

%Let us now ask the question: 
%what is the physical meaning of 
%the Nambu-Poisson structure? 

The M5-brane action obtained from the BL action 
with the Nambu-Poisson algebra 
should be interpreted as the M5-brane theory 
in a large $C$-field background. 
In Sec.\ref{SUSY}, we find this interpretation 
to be consistent with the properties of the supersymmetry. 
Furthermore, in Sec.\ref{M5D4}, 
the noncommutative D4-brane action obtained via 
a double dimensional reduction from the M5 theory 
has $g\sim \theta$. 
As it is well known that for a D4-brane 
the noncommutativity parameter $\theta$ 
is given by $B^{-1}$ in the large $B$-field background, 
we deduce that
(with the specific normalization of $C$-field
such that the self-dual field strength
becomes $H=db+C$)
\begin{equation} 
g \sim C^{-1}
\end{equation}
in the large $C$-field background for our M5-brane theory.

In \cite{Ho:2008nn},
an analogy was made between 
the Nambu-Poisson structure on M5-brane 
and the Poisson structure on D-branes. 
For a D-brane in a constant $B$-field background, 
the effective D-brane theory 
is best described as a noncommutative field theory. 
In the limit of both large and small $B$-field background,
the noncommutativity is small 
and the commutator can be approximated 
by a Poisson bracket,  
and the Poisson structure is determined 
by the two-form $B$-field background. 
More precisely, the gauge-invariant quantity 
which should be used to specify the background is 
${\cal F} = B + F$, 
where $F=dA$ is the field strength 
of a gauge field $A$ in the D-brane world-volume theory. 
We can fix the gauge so that 
the background value of ${\cal F} = B$. 
Then a nontrivial configuration of the gauge field $A$ 
corresponds to a change of ${\cal F}$, 
and thus a change of the Poisson structure. 

%ima:I replaced A -> b
%Based on this analogy, 
%we propose that the physical meaning of 
%the Nambu-Poisson structure 
%is a $C$-field background. 
The invariant self-dual 3-form field strength on 
the M5-brane is $H = C + db$, 
where $b$ is the 2-form gauge potential 
on the M5-brane. 
The Nambu-Poisson structure 
determined by a given background $H = C$ 
is therefore deformed by turning on $b$, 
while a gauge transformations of $b$ 
preserves $H$ and thus the Nambu-Poisson structure.

In the case of D-branes in $B$-field background, 
there are several different ways to verify the connection 
between $B$-field background and the noncommutativity. 
One way is to quantize an open string ending on a D-brane 
and check that the endpoint coordinates 
obey a commutation relation determined by the $B$-field 
background \cite{ChuHo}. 
Or one can compute open string scattering amplitudes 
\cite{Schomerus}.
Another way \cite{SeibergWitten}
is to find the Seiberg-Witten map 
which maps the commutative field $A$ to 
the noncommutative field $\hat{A}$, 
and then check that the (commutative) D-brane field theory 
with $B$-field background explicitly turned on 
is approximately the same as 
the noncommutative field theory without 
explicit $B$-field background. 

Quantization of open membranes in 
the large $C$-field background has been 
extensively studied in the literature \cite{membraneC,Matsuo:2000fh}. 
However, quantization is by its nature 
associated with a Poisson structure, 
and thus the appearance of a Nambu-Poisson structure 
can not be manifest. 
On the other hand, in \cite{Ho:2007vk}, 
the scattering amplitudes of open membranes 
were studied in a large $C$-field background, 
and the result indicated that indeed 
the $C$-field background induces 
a Nambu-Poisson structure. 

As a further support of our interpretation of 
the Nambu-Poisson structure as an effect of 
the $C$-field background, 
in the previous section we
found the Seiberg-Witten map 
which matches the gauge transformation
of ordinary M5-brane theory 
with the deformed gauge transformation (\ref{gt1}-\ref{gt4}) 
supposedly corresponding to a $C$-field background. 
%It remains to be seen whether this will allow us to connect 
%the M5-brane action with explicit $C$-field dependence 
%\cite{rM51,rM52} 
%to our M5-brane model with a Nambu-Poisson structure.

%%%%%%%%%%%%%%%%%%%%%%%%%%%%%%%%%%%%%%%%%%%%%%%%%%%%%%%%%%%%%
\section{Further remarks}
\paragraph{Quantization of the coefficients of Chern-Simons term}

In section \ref{3to6}
we showed that if the internal space is ${\cal N}={\bf R}^3$
the model has no coupling constant at all.
What happens when ${\cal N}$ is a non-trivial space?
In such a case there may not be any simple way to remove the coupling constant
by re-scaling of fields.
It is an interesting problem to clarify constraints
imposed on this coupling constant.
For the case when the 3-algebra is taken to be $\mathcal{A}_4$, 
Bagger and Lambert \cite{BaggerLambert}
have shown that the eigenvalues of the structure constant
$f^{abcd}\sim \epsilon^{abcd}$ must be quantized as
$\lambda=\pi/k$ for $k=1,2,3\cdots$.
It implies that in BLG model there are no tunable continuous parameters
in the theory.  In this paper, we have used Lie 3-algebras which
has an infinite number of generators.  One may wonder if
we might have a similar constraint for the structure constant,
especially if the internal space $\mathcal{N}$ is a compact space.
For $\mathcal{N}=T^3$, for example, the generators
are labeled by $\vec n\in\mathbf{Z}^3$
and the structure constant
becomes \cite{Ho:2008nn}
\ba\label{torusalg}
f^{\vec n_1 \vec n_2 \vec n_3 \vec n_4}= \frac{\alpha}{V}\,\vec n_1\cdot
(\vec n_2\times \vec n_3) \delta_{\vec n_1+\vec n_2+\vec n_3+\vec n_4,0}\,,
\ea
where $V$ is the volume of $T^3$ and $\alpha$ is a constant.
It is known that for  any Lie 3-algebra with finite number of generators
and positive invariant metric can be reduced to the direct sum
of $\mathcal{A}_4$ \cite{positivedefinitemetric}.
Here we have a consistent Lie 3-algebra with positive definite metric
while the number of generators is infinite. A natural question is
whether our algebra such as (\ref{torusalg}) can be understood as
the direct (and infinite) sum of $\mathcal{A}_4$.  If this is the case,
we need to have a similar quantization condition for the structure
constant.  It will be very interesting if such quantization of parameter
exists in the compact internal space $\mathcal{N}$.

\paragraph{Global structure of internal space}
The classification of possible internal manifold $\mathcal{N}$
is another challenging issue.
What is required in this paper is that (i) $\mathcal{N}$ is covered by
patches with
local coordinates $y^{\dot\mu}$ and (ii) on the intersection of
different patches
the local coordinates are related with each other 
by the volume-preserving
diffeomorphism.  $T^3$ is an obvious example of manifold with such structure.
In order to understand the relation between M2 and M5,
the mathematical classification $\mathcal{N}$ will be indispensable.

\paragraph{Multiple/long M5-brane}
In our paper, we construct a single M5-brane action from the BLG
model.  One of the most challenging issue is how to construct the action of
{\em multiple} M5-branes.
For that purpose, we need to construct a set of generators
$T^A \chi^a(y)$, where $T^A$ ($A=1,\cdots,d$) are the generators
of an internal algebra and $\chi^a(y)$ is the basis
of functions on $\mathcal{N}$.
However, as far as we try, it seems difficult to find 3-algebras
of this form which satisfies the fundamental identity.

One idea to understand the nature of this problem is to consider
the multiple cover of $\mathcal{N}$.  Let us take the simplest
example $T^3$ and take all of its radius to $2\pi$. for simplicity.
Then the basis of functions is of the form
$\exp(i\sum_{\dot\mu=\dot1}^{\dot3} n_{\dot \mu} y^{\dot\mu})$
where $n_{\dot\mu}$ is integer.
Suppose one takes the double cover in $y^{\dot1}$ direction.
Then it may be possible to take $n_{\dot 1}$ to be half integer.
So we have two sets of generators, one $\chi^{\vec n}$ for
$n_{\dot 1}\in \mathbf{Z}$ and the other $\chi^{\vec n}$ for
$n_{\dot 1}\in \mathbf{Z}+1/2$.  We write the former generators
as $T^{\vec n}$ and the latter as $S^{\vec n}$.  It is then elementary
to show that
\ba
\{T,T,T\}\sim T,\quad
\{T,T,S\}\sim S,\quad
\{T,S,S\}\sim T,\quad
\{S,S,S\}\sim S\,.
\ea
So the 3-algebra of original $T^3$ is contained in the algebra
of covering space as a subalgebra.
It is not difficult to show that similar effect occurs in general.
Namely let us denote the 3-algebra associated with 3-manifold
$\mathcal{N}$ as $\mathcal{A}_{\mathcal{N}}$ and
let $\tilde\mathcal{N}$ be a covering space of $\mathcal{N}$.
Then $\mathcal{A}_{\mathcal{N}}$ becomes a subalgebra
of  $\mathcal{A}_{\tilde\mathcal{N}}$.
Since $\mathcal{A}_{\tilde\mathcal{N}}$ is not the direct
product of $\mathcal{A}_{\tilde\mathcal{N}}$ with finite Lie 3-algebra
as above, $\mathcal{A}_{\tilde\mathcal{N}}$ does not describe
multiple M5 but it describes
long M5 which wraps $\mathcal{N}$ several times.
Such a connection, however, may be helpful
to understand the multiple M5 in the future.

\paragraph{Vortex string and volume-preserving diffeomorphism}
As we commented, in our construction of M5-brane action, we do not
need the metric on $\mathcal{N}$ but only its volume form,
or in other words, the 3-form flux $C$ on it.  Our computation
further implied that it is natural to assume that
there is a very large 3-form flux $C$ on the M5 world-volume.
This set-up reminds us of the open membrane in large
$C$ flux.  Since we can neglect the Nambu-Goto part
(which contains the metric), the action becomes
that of the topological membrane \cite{topmeb},
\ba
S\sim \int C_{\mu\nu\rho} dX^\mu \wedge dX^\nu \wedge dX^\rho\,.
\ea
When this membrane has the boundary on M5, this topological
action gives
\ba
S\sim \int C_{\mu\nu\rho} X^\mu dX^\nu \wedge  dX^\rho\,.
\ea
It gives an action for the string which describes the
boundary of the open membrane.  When the target space has 3 dimensions
and $C\sim \epsilon^{\dot\mu\dot\nu\dot\lambda}$, this action
is identical to the kinetic term of the vortex string \cite{Lund:1976ze},
which was found long ago.
In the supermembrane context it was studied in
\cite{membraneC,Matsuo:2000fh, Ho:2007vk}.
In particular it was found that it can be equipped with
the Poisson structure with the constraint associated with
the diffeomorphism which defines the volume-preserving
diffeomorphism naturally \cite{Matsuo:2000fh},
\ba
&&\delta X^{\dot\mu}=\{ X^{\dot\mu}, \omega(f,g)\}_D= v^{\dot\mu}(X) +\cdots
\,,\\
&& v^{\dot\mu}=\epsilon^{\dot\mu\dot\nu\dot\lambda} \partial_{\dot \nu}
f \partial_{\dot\lambda} g\,,\quad \partial_{\dot\mu}v^{\dot \mu}=0 \,,\\
&& \omega(f,g):= \int d\sigma f(X) dg(X)\,.
\ea
Here $\{\,,\,\}_D$ is the Dirac bracket associated with
the kinetic term and $\cdots$ in the first line describe the
extra variation along the world-sheet which can be absorbed
by the reparametrization of the world-sheet.
In Bagger-Lambert theory, the gauge parameter has an unusual
feature that it has two index $\Lambda_{ab}$.  In this picture,
this structure is naturally interpreted as a result of the fact
that for the string we can introduce two functions $f,g$ to define the
generators on the world-sheet.
We hope that this connection with the vortex string would give
a new insight into the BLG model.

\section*{Acknowledgment}

We appreciate partial financial support from
Japan-Taiwan Joint Research Program
provided by Interchange Association (Japan)
by which this collaboration is made possible.

The authors thank Kazuyuki Furuuchi, Darren Sheng-Yu Shih, 
and Wen-Yu Wen for helpful discussions. 
P.-M. H. is grateful to Anna Lee for assistance in many ways. 
The work of P.-M. H. is supported in part by
the National Science Council,
the National Center for Theoretical Sciences, 
and the LeCosPA Center at National Taiwan University. 
Y. M. is partially supported by
Grant-in-Aid (\#20540253) from the Japan
Ministry of Education, Culture, Sports,
Science and Technology.
Y.I. is partially supported by
a Grant-in-Aid for Young Scientists (B) (\#19740122) from the Japan
Ministry of Education, Culture, Sports,
Science and Technology.

\appendix
\section{Derivation of some equations}

In this appendix we derive some equations used in the main text.
We first derive the commutation relation (\ref{comm3}).
By using the explicit form of the
covariant derivative in (\ref{dmu}),
we obtain
\begin{equation}
[{\cal D}_\mu,{\cal D}_\nu]\Phi
=-gF_{\mu\nu}^{\dot\kappa}\partial_{\dot\kappa}\Phi,
\label{noncovf}
\end{equation}
where the explicit form of $F_{\mu\nu}^{\dot\kappa}$
in terms of the potential is
\begin{equation}
F_{\mu\nu}^{\dot\kappa}=\epsilon^{\dot\kappa\dot\mu\dot\nu}
\partial_\mu\partial_{\dot\mu}b_{\nu\dot\nu}
-g\epsilon^{\dot\mu\dot\nu\dot\rho}
\partial_{\dot\mu}b_{\mu\dot\nu}
\epsilon^{\dot\kappa\dot\lambda\dot\tau}
\partial_{\dot\rho}
\partial_{\dot\lambda}b_{\nu\dot\tau}
-(\mu\leftrightarrow\nu).
\end{equation}
Because the (non-covariant) derivative appears
on the right hand side,
$F_{\mu\nu}^{\dot\kappa}$ defined by
(\ref{noncovf}) is not covariant.
We can define the covariantized $F$ by
\begin{equation}
[{\cal D}_\mu,{\cal D}_\nu]\Phi
=-g{\cal F}_{\mu\nu}^{\dot\kappa}{\cal D}_{\dot\kappa}\Phi.
\label{covf}
\end{equation}
These two fields are related by
$F_{\mu\nu}^{\dot\kappa}\partial_{\dot\kappa}\Phi
={\cal F}_{\mu\nu}^{\dot\kappa}{\cal D}_{\dot\kappa}\Phi$,
and by substituting $\Phi=X^{\dot\mu}$ into this relation
and using 
\begin{equation}
g{\cal D}_{\dot\mu}X^{\dot\sigma}
=V\delta_{\dot\mu}^{\dot\sigma},
\label{gdelx}
\end{equation}
we obtain
\begin{equation}
V{\cal F}_{\mu\nu}^{\dot\kappa}
=gF_{\mu\nu}^{\dot\lambda}\partial_{\dot\lambda}X^{\dot\kappa}.
\label{calfandf}
\end{equation}
${\cal F}$ can be expressed as the covariant derivative
of the field strength ${\cal H}$.
\begin{equation}
V{\cal F}_{\mu\nu}^{\dot\kappa}
=g{\cal F}_{\mu\nu}^{\dot\lambda}{\cal D}_{\dot\lambda}X^{\dot\kappa}
=\epsilon_{\mu\nu\lambda}
    \epsilon^{\lambda\rho\sigma}{\cal D}_\rho{\cal D}_\sigma X^{\dot\kappa}
=\epsilon_{\mu\nu\lambda}
{\cal D}_\rho
\wt{\cal H}^{\rho\lambda\dot\kappa}.
\label{fandh}
\end{equation}
In the first step we used the relation
(\ref{gdelx}).
Substituting this into (\ref{covf}),
we obtain (\ref{comm3}).

Next, let us consider the equations of motion
of the gauge fields $b_{\dot\mu\dot\nu}$ and $b_{\mu\dot\nu}$.
For a variation of $b_{\dot\mu\dot\nu}$,
we have the following variations of the action.
\begin{equation}
\delta S_X
=\int d^3x\langle\delta b^{\dot\mu}{\cal D}_\mu{\cal D}_\mu X^{\dot\mu}\rangle
=\frac{1}{2}\int d^3x\langle \delta b^{\dot\mu}
\epsilon^{\dot\mu\dot\rho\dot\sigma}{\cal D}_\mu  {\cal H}_{\mu\dot\rho\dot\sigma}\rangle
\end{equation}
\begin{equation}
\delta S_{\rm pot}
=\frac{g^4}{2}\int d^3x\langle
\delta b^{\dot\mu}\{X^I,X^J,\{X^I,X^J,X^{\dot\mu}\}\}
\rangle,
\end{equation}
\begin{equation}
\delta S_{\rm int}
=\frac{ig^2}{2}\int d^3x
\langle\ol\Psi,\Gamma_{\dot\mu J}\{\delta b^{\dot\mu},X^J,\Psi\}\rangle
=-\frac{ig^2}{2}
\int d^3x\langle\delta b^{\dot\mu}\{\ol\Psi\Gamma_{\dot\mu J},X^J,\Psi\}\rangle,
\end{equation}
and the equation of motion is
\begin{eqnarray}
0
&=&
\frac{1}{2}
\epsilon^{\dot\mu\dot\rho\dot\sigma}{\cal D}_\mu
{\cal H}_{\mu\dot\rho\dot\sigma}
+\frac{g^4}{2}\{X^I,X^J,\{X^I,X^J,X^{\dot\mu}\}\}
-\frac{ig^2}{2}\langle\delta b^{\dot\mu}\{\ol\Psi\Gamma_{\dot\mu J},X^J,\Psi\}\rangle
\nonumber\\
&=&
\frac{1}{2}
\epsilon^{\dot\mu\dot\rho\dot\sigma}{\cal D}_\mu
{\cal H}_{\mu\dot\rho\dot\sigma}
+{\cal D}_{\dot\mu}{\cal H}_{\dot1\dot2\dot3}
+g^2\epsilon^{\dot\rho\dot\mu\dot\tau}
\{X^i,X^{\dot\rho},{\cal D}_{\dot\tau}X^i\}
\nonumber\\&&
+\frac{g^4}{2}\{X^i,X^j,\{X^i,X^j,X^{\dot\mu}\}\}
-\frac{ig^2}{2}\{\ol\Psi\Gamma_{\dot\mu J},X^J,\Psi\},
\end{eqnarray}
or, equivalently,
\begin{equation}
{\cal D}_\mu{\cal H}^{\mu\dot\rho\dot\sigma}
+{\cal D}_{\dot\mu}{\cal H}^{\dot\mu\dot\rho\dot\sigma}
=gJ^{\dot\rho\dot\sigma},
\end{equation}
where the current is given in the text.

For the variation of the gauge field $b_{\lambda\dot\mu}$,
we obtain
\begin{eqnarray}
\delta S_X
&=&
-g\int d^3x\langle
\delta b_{\lambda\dot\mu}\{X^I,{\cal D}_{\lambda}X^I,y^{\dot\mu}\}\rangle
,\\
\delta S_\Psi
&=&-\frac{ig}{2}\int d^3x\langle
\ol\Psi\Gamma^\lambda\{\delta b_{\lambda\dot\mu},y^{\dot\mu},\Psi\}\rangle
=-\frac{ig}{2}
\int d^3x\langle
\delta b_{\lambda\dot\mu}\{\ol\Psi\Gamma^\lambda,\Psi,y^{\dot\mu}\}
\rangle,\\
\delta S_{\rm CS}
&=&-\frac{1}{2}\int d^3x\langle \epsilon^{\lambda\mu\nu}
\delta b_{\lambda\dot\mu}F_{\mu\nu}^{\dot\mu}\rangle.
\end{eqnarray}
The equation of motion for $b_{\lambda\dot\mu}$ is
\begin{equation}
\frac{1}{2}\epsilon^{\lambda\mu\nu}F_{\mu\nu}^{\dot\mu}
+g\{X^I,{\cal D}_\lambda X^I,y^{\dot\mu}\}
+\frac{ig}{2}\{\ol\Psi\Gamma^\lambda,\Psi,y^{\dot\mu}\}
=0.
\end{equation}
This is not covariant, but we can covariantize this by multiplying
$g\partial_{\dot\mu}X^{\dot\nu}$.
\begin{equation}
\frac{V}{2}\epsilon^{\lambda\mu\nu}{\cal F}_{\mu\nu}^{\dot\mu}
+g^2\{X^I,{\cal D}_\lambda X^I,X^{\dot\mu}\}
+\frac{ig^2}{2}\{\ol\Psi\Gamma^\lambda,\Psi,X^{\dot\mu}\}
=0.
\end{equation}
By using (\ref{h12def}) and (\ref{fandh}),
we can rewrite this equation of motion as follows:
\begin{equation}
\wt{\cal D}_\rho{\cal H}^{\rho\lambda\dot\mu}
+{\cal D}_{\dot\kappa}{\cal H}^{\dot\kappa\lambda\dot\mu}
=
gJ^{\lambda\dot\mu}
\end{equation}

%%%%%%%%%%%%%%%%%%%%%%%%%%%%%%%%%%%%%%%%%%%%%%%%%%%%%%%%%

The Bianchi identity (\ref{maxwell3}) is obtained by
substituting $\Phi=X^{\dot\mu}$ to the commutation relation
(\ref{comm2}).
By using the definition of the field strength ${\cal H}$,
we can rewrite the left hand side as
\begin{eqnarray}
[{\cal D}_\lambda,{\cal D}_{\dot\lambda}]X^{\dot\mu}
&=&\delta_{\dot\lambda}^{\dot\mu}{\cal D}_\lambda{\cal H}_{\dot1\dot2\dot3}
-\frac{1}{2}\epsilon^{\dot\mu\dot\rho\dot\sigma}{\cal D}_{\dot\lambda}
{\cal H}_{\lambda\dot\rho\dot\sigma},
\end{eqnarray}
and the right hand side becomes
\begin{equation}
g^2\{{\cal H}_{\lambda\dot\nu\dot\lambda},X^{\dot\nu},X^{\dot\mu}\}
=\epsilon^{\dot\nu\dot\mu\dot\kappa}{\cal D}_{\dot\kappa}
{\cal H}_{\lambda\dot\nu\dot\lambda}.
\end{equation}
Combining these, we obtain the Bianchi identity
\begin{equation}
{\cal D}_\lambda{\cal H}_{\dot\lambda\dot\rho\dot\sigma}
-{\cal D}_{\dot\lambda}
{\cal H}_{\lambda\dot\rho\dot\sigma}
-{\cal D}_{\dot\rho}{\cal H}_{\lambda\dot\sigma\dot\lambda}
-{\cal D}_{\dot\sigma}{\cal H}_{\lambda\dot\lambda\dot\rho}=0.
\end{equation}
This is equivalent to (\ref{maxwell3})

\end{document}